\documentclass[a4paper,12pt]{article}

\usepackage{pstricks}

\usepackage[sumlimits,intlimits,namelimits]{amsmath} 
\usepackage{dsfont}
\usepackage{slashed}

\usepackage[english]{babel}

\usepackage{graphicx}
\usepackage{subfig,caption}
\usepackage{amssymb}
\usepackage{amsmath}

\usepackage{cite}

\usepackage{geometry}
\newcommand{\op}{{\cal O}}

\newcommand{\nn}{\nonumber}

\newcommand{\av}[1]{\langle #1 \rangle}

\newcommand{\sla}[1]{#1 \!\!\!/}

\newcommand{\skipp}[1]{}
\newcommand{\A}{{\cal A}}

\makeatletter
\def\fmslash{\@ifnextchar[{\fmsl@sh}{\fmsl@sh[0mu]}}
\def\fmsl@sh[#1]#2{%
  \mathchoice
    {\@fmsl@sh\displaystyle{#1}{#2}}%
    {\@fmsl@sh\textstyle{#1}{#2}}%
    {\@fmsl@sh\scriptstyle{#1}{#2}}%
    {\@fmsl@sh\scriptscriptstyle{#1}{#2}}}
\def\@fmsl@sh#1#2#3{\m@th\ooalign{$\hfil#1\mkern#2/\hfil$\crcr$#1#3$}}
\makeatother

\numberwithin{equation}{section}
\begin{document}
\begin{titlepage}
\begin{flushright}
SI-HEP-2014-28\\
QFET-2015-19 \\[0.2cm]
\end{flushright}

\vspace{1.2cm}
\begin{center}
{\Large\bf 
Three-Body Non-Leptonic \boldmath $B$ \unboldmath Decays \\[2mm]  and QCD Factorization}
\end{center}

\vspace{0.5cm}
\begin{center}
 {\sf Susanne Kr\"ankl, Thomas Mannel and Javier Virto} \\[0.2cm]
{\it Theoretische Elementarteilchenphysik, Naturwiss.- techn. Fakult\"at, \\
Universit\"at Siegen, 57068 Siegen, Germany}
\end{center}

\vspace{0.8cm}
\begin{abstract}
\vspace{0.2cm}\noindent
We extend the framework of QCD factorization to non-leptonic $B$ decays into three light mesons,
taking as an example the decay $B^+ \to \pi^+ \pi^+ \pi^-$.
We discuss the factorization properties of this decay in different regions of phase space.
We argue that, in the limit of very large $b$-quark mass, the central region of the Dalitz plot can be described in terms of the $B \to \pi$ form factor 
and the $B$ and $\pi$ light-cone distribution amplitudes.
The edges of the Dalitz plot, on the other hand, require different non-perturbative input: the $B \to \pi \pi$ form factor and the two-pion distribution
amplitude. We present the set-up for both regions to leading order in both $\alpha_s$ and $\Lambda_{\rm QCD} / m_b$ and discuss how well the two descriptions merge.
We argue that for realistic $B$-meson masses there is no perturbative center in the Dalitz plot, but that a systematic description might be possible in the context of
two-pion states. As an example, we estimate the $B\to\rho\pi$ branching fraction beyond the quasi-particle approximation. We also discuss the prospects for studies of three-body
and quasi-two-body non-leptonic $B$ decays from QCD. 
\end{abstract}

\end{titlepage}

\newpage
\pagenumbering{arabic}
\section{Introduction}
Three-body non-leptonic decays of heavy mesons constitute a large portion of the 
branching fraction. For $B$ mesons, three-body non-leptonic branching ratios and CP asymmetries have been measured
for a large number of channels, most notably by BaBar, Belle and LHCb~\cite{0902.2051,1301.3186,1308.0740,1310.4740,1408.5373,1501.00705},
and more is  expected to come from LHC run 2 and from Belle II~\cite{1002.5012}.
On the theory side, three-body non-leptonic $B$ decays are interesting for several phenomenological 
applications, such as the study of CP violation and the extraction of the CKM angles $\alpha$ and $\gamma$ (see e.g. \cite{Snyder:1993mx,0210433}). 
While in most cases there is a dominance of quasi two-body final states, in some decay channels the 
contributions from non-resonant three-particle states seems to be rather large~\cite{1412.7515}. 
The study of the interference pattern of the resonances in Dalitz plots is a well established method to 
determine CP asymmetries~\cite{9809262,0905.4233}, while further information can be inferred
on strong resonances, such as masses,  widths and quantum numbers~\cite{0910.0454}.

There are, however, two obvious problems in the quasi-two-body interpretation of resonant effects in multi-body decays,
one practical and one conceptual. From the practical point of view, any parametrization of resonant structures is model dependent, as no
universal line-shape for strong resonances is accurate, especially for broad states. On the conceptual level, the
mere separation of resonant and non-resonant contributions is not clear-cut, most prominently in the case of non-leptonic decays where non-factorizable effects exist.    

In the case of two-body non-leptonic $B$ decays, the heavy-quark limit has been exploited systematically in the
context of QCD factorization \cite{9905312,0006124,0104110,0308039,0603239} or Soft-Collinear
Effective Theory (SCET) \cite{0011336,0206152,0401188,0510241,0601214}, where the matrix elements
factorize into a convolution of perturbative hard kernels, form factors and meson distribution amplitudes on the
light cone. Corrections to factorization arise at subleading orders in the heavy-quark / large-energy expansion, and
remain a source of uncertainty which is difficult to estimate.
Some potentially important non-factorizable effects might be related to
nearly on-shell intermediate states, such as charm-loops or rescattering effects from light mesons.
Phenomenological investigations of such effects have limited potential, mainly because the kinematics of two-body
decays is fixed.
On the other hand, three-body decays have at their disposal a wide phase space where the
energy dependence of such effects can be studied, with the potential of providing a deeper understanding of
factorization and hadronic effects in $B$ decays.     

It is fair to say that 
the theoretical description of three-body $B$ decays is still in the stage of modeling.
Common methods reflecting the state of the art are the isobar model~\cite{Sternheimer:1961zz,Herndon:1973yn} and the K-matrix formalism~\cite{Chung:1995dx}. 
In these approaches, resonances are modeled and the non-resonant contributions are often described by an 
empirical distribution in order to reproduce the full range of the phase space \cite{0412066}. 
In the context of factorization, in Refs.~\cite{0209164,0506268,0704.1049} the matrix elements
were factorized naively and the resulting local correlators were computed in the framework of Heavy-Meson
Chiral Perturbation Theory (HMChPT), but no attempt was made to address the breakdown of factorization or
HMChPT in the respective regions of phase space where they are not expected to apply.
Other recent work relying on pQCD~\cite{0209043}, seems to reproduce experimental values for CP asymmetries integrated in certain regions of phase space~\cite{1402.5280}.
However, if the conceptual issues regarding the  pQCD approach~\cite{0109260,0901.2965} cannot be resolved, its predictive power remains limited.
In the future, novel model-independent approaches 
that directly access CP violation (such as the Miranda procedure~\cite{0905.4233,1205.3036})
or methods based on flavor symmetries~\cite{1011.4972,1106.2511}   
could become interesting; however, for a quantitative description of the differential Dalitz distributions including
amplitude phase information, a QCD-based approach is unavoidable.


In the present letter we take a step in this direction, and study the factorization properties of 
charmless three-body and the corresponding quasi-two body $B$ decays\footnote{The main ideas developed here have been discussed qualitatively by M.~Beneke~\cite{talkMB} and I.~Stewart~\cite{talkIS}.}. 
For that purpose, we focus specifically on the decay $B^+ \rightarrow\pi^+ \pi^- \pi^+$, assuming that the $b$-quark
mass is large enough.
We start by identifying the different regions in the Dalitz plot where the well-established factorization properties of two-body
decays apply to the three-body case. In the heavy-mass limit, we discuss how to compute the central region of the Dalitz plot
as well as its edges. We will see that the methods and the theoretical inputs are different in the different regions:
while the center can be described in terms of regular from factors and pion distributions, the description at the
edges requires introducing generalized versions of these hadronic matrix elements. Generalized form factors and
distribution amplitudes have been already studied, the former in the context of semileptonic $B$ decays \cite{1310.6660},
and the latter in connection with two-meson electro-production \cite{Grozin:1983tt,9805380}
or semileptonic $\tau$ decays \cite{Kuhn:1990ad}
(see e.g. \cite{Baier:1985pd,Grozin:1986at,9809483,0003233,0307382,0805.3773,0807.4883,0908.3589,1006.5313,1203.6839,1301.6973}).
As an application, we consider the $B^+\to \rho^0 \pi^+$ branching ratio by integrating the differential rate around
the $\rho$ resonance. Finally, we discuss how both descriptions merge to describe the full Dalitz plot, and what we can
expect for realistic $b$-quark masses.

\section{Identifying regions in the Dalitz plot} 

We consider the decay $B^+ \rightarrow\pi^+ \pi^- \pi^+$, and define the external momenta as:
\begin{equation}
 B^+ (p) \to \pi^+(k_1) \, \, \pi^- (k_2) \, \, \pi^+(k_3)  \quad \mbox{with}  \quad p = k_1 + k_2 + k_3
 \ \ \text{and}\ \ E_1^\text{CM}\le E_3^\text{CM}\ , 
\end{equation}
where CM refers to the $B$-meson rest frame. We neglect pion masses in the kinematics, such that:
\begin{equation} 
p^2=m_B^2 \ \ , \quad k_i^2 = 0 \ , \quad
s_{ij} \equiv \frac{(k_i + k_j)^2}{m_B^2} = \frac{2 k_i\cdot k_j}{m_B^2}  \, \quad (i \neq j) \ .
\end{equation} 

\begin{figure}
\centerline{
\includegraphics[scale=0.75]{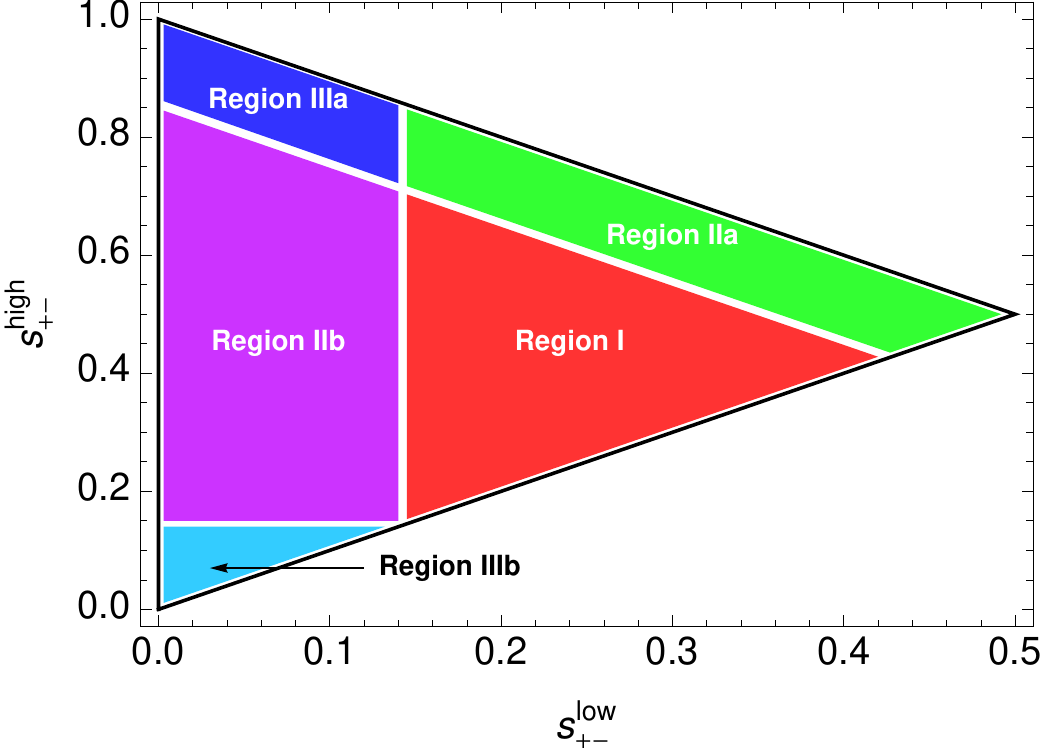}\hspace{8mm}
\includegraphics[scale=0.75]{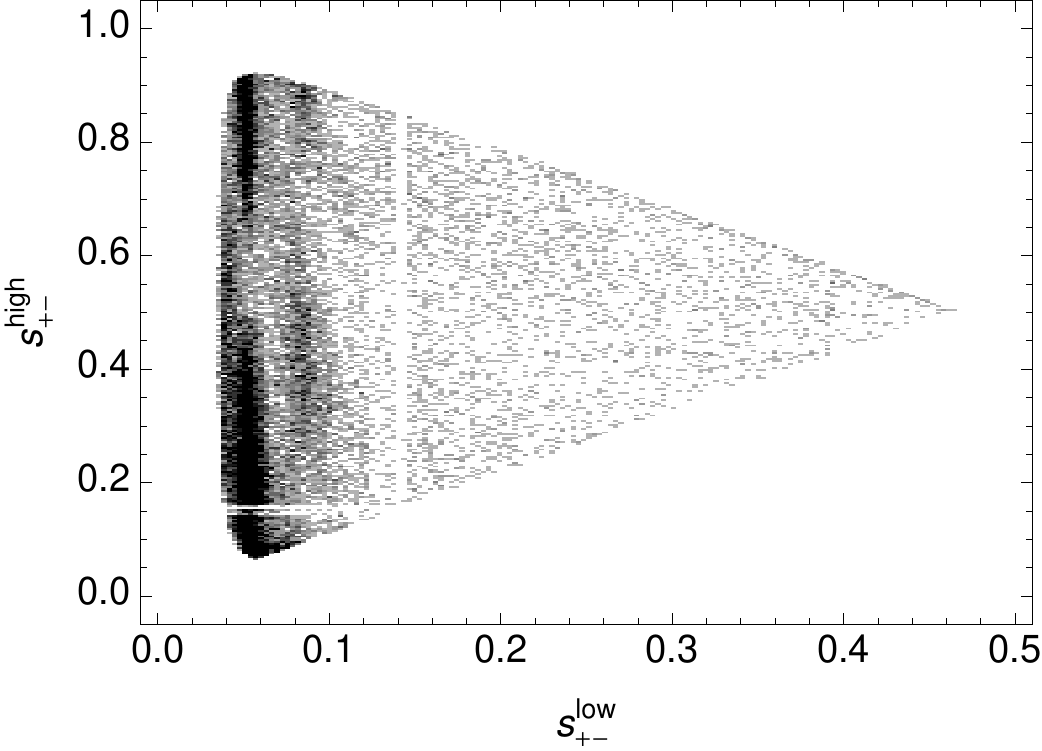}
}
\caption{Left: The physical kinematical region in the plane of two independent momentum invariants $s_{+-}^\text{low}$, $s_{+-}^\text{high}$ (Dalitz plot), divided into the different regions with special kinematical configurations:
I - Mercedes Star configuration, IIa,IIb - Two collinear pions, IIIa,b - One soft pion. Right: Dalitz plot distribution
for $B^+\to \pi^+\pi^-\pi^+$ from Ref.~\cite{1408.5373}}
\label{Dalitz} 
\end{figure}

The kinematics of the three-body decay is completely determined by two of the three kinematic invariants
$s_{12} \equiv s_{+-}^\text{low}$, $s_{13} \equiv s_{++}$ and $s_{23}\equiv s_{+-}^\text{high}$, which (in the massless limit) satisfy $s_{12}+s_{13}+s_{23} = 1$ and $0\le s_{ij}\le 1$.
The physical kinematical region in the plane of two invariants (the Dalitz plot) is given in this case by a triangle
(see Fig.~\ref{Dalitz}).

\bigskip

We distinguish three special kinematical configurations:
\begin{enumerate}
\item[\bf I.]  {\bf ``Mercedes Star'' configuration:} This corresponds to the central region of the Dalitz plot,
where all the invariant masses are roughly the same and of order of $m_B$:
\begin{equation}
\textbf{Region I : }\quad s_{++} \sim s_{+-}^\text{low} \sim s_{+-}^\text{high} \sim 1/3
\end{equation} 
corresponding to the kinematical situation where all three pions have a large energy in the $B$-meson rest frame and 
none of the pions moves collinearly to any other.
\item[\bf II.] {\bf Collinear decay products:} This corresponds to regions of the Dalitz plot where one invariant mass is small and
the other two are large.  The kinematic configuration is such that two pions are collinear, generating a small invariant mass  recoiling against the third pion. In our case there are two such regions:
\begin{equation}
\textbf{Region IIa : }\quad s_{++}  \sim 0 \, ,\quad s_{+-}^\text{low} \sim s_{+-}^\text{high} \sim 1/2
\end{equation}
which is the region where the two $\pi^+$ move collinearly, recoiling against the $\pi^-$, and
\begin{equation}
\textbf{Region IIb : }\quad s_{+-}^\text{low}  \sim 0 \, ,\quad s_{++}\sim s_{+-}^\text{high} \sim 1/2
\end{equation}
where the $\pi^-$ and one $\pi^+$ move collinearly, recoiling against the second $\pi^+$.

\item[\bf III.] {\bf One soft decay product:} The regions of the Dalitz plot where two invariant masses are small and one is large correspond to kinematical configurations where one pion is soft and the other two are fast and back-to-back. In our case there are two such regions:
\begin{equation}
\textbf{Region IIIa : }\quad s_{++}  \sim  s_{+-}^\text{low}\sim 0\, ,\quad s_{+-}^\text{high} \sim 1
\end{equation}
which is the region where one $\pi^+$ is soft, and
\begin{equation}
\textbf{Region IIIb : }\quad s_{+-}^\text{high}  \sim  s_{+-}^\text{low}\sim 0\, ,\quad s_{++} \sim 1
\end{equation}
where the $\pi^-$ is soft.

\end{enumerate}

The different regions are shown in Fig.~\ref{Dalitz}. For a very heavy $B$ meson, region I is dominant, since the condition $m_B^2 s_{ij} \gg \Lambda_{\rm QCD}^2$ 
is satisfied in most of the Dalitz plot. The edges of the Dalitz plot, corresponding to collinear and soft configurations, will be small. 
However, in the edges, all the resonances show up, corresponding to quasi-two particle decays. The masses $m_R$ of 
these resonances do not scale with the heavy $b$ quark mass and hence the width of regions II and III scale as $m_R / m_B$, 
showing the dominance of region I in the infinite mass limit. 

\bigskip

In the following we propose to perform a QCD factorization calculation in region I of the Dalitz plot in terms of the pion
and $B$-meson light-cone distributions and the $B \to \pi$ form factor.
The presence of resonances in region IIb will be signaled  by a singular behavior of the factorized amplitudes of the
form $1/s_{+-}^\text{low}$.
A proper treatment this region requires to set up a different calculational method, which requires a different form of
QCD factorization.  In this case, new non-perturbative quantities need to be defined: a light-cone distribution for two pions,
and a form factor for the $B \to \pi \pi $ transition. Finally, it is important to check that the two calculations match properly
in order to obtain a complete description of all regions in the Dalitz plot.

The region IIa contains no resonances (corresponding to the $\pi^+\pi^+$ channel), and will see that the factorized
amplitudes are regular as $s_{++} \to 0$. In this case the ``perturbative" result should provide a good description of the rate
when integrated (or smeared) over a suitable interval, in the sense of parton-hadron duality.

\section{The central region of the Dalitz plot}
\label{sec:MS}
\begin{figure}
\begin{center}
\includegraphics[width=6cm]{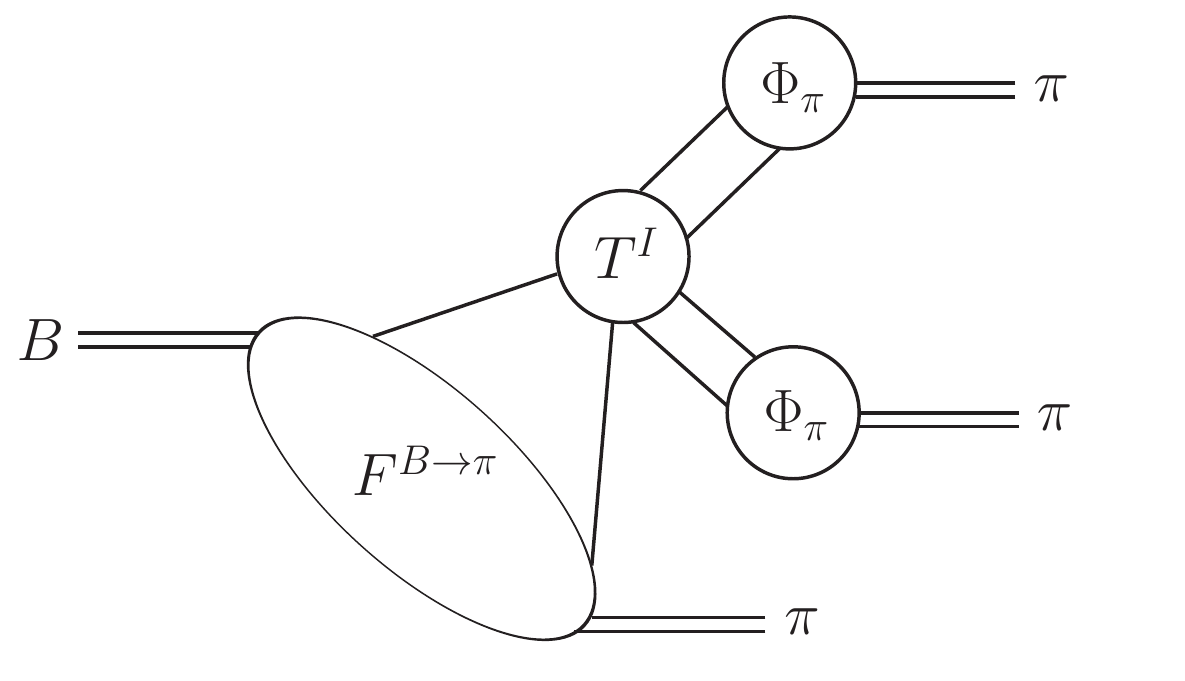} 
\hspace{1cm}
\raisebox{15mm}{+}
\hspace{1cm}
\includegraphics[width=6cm]{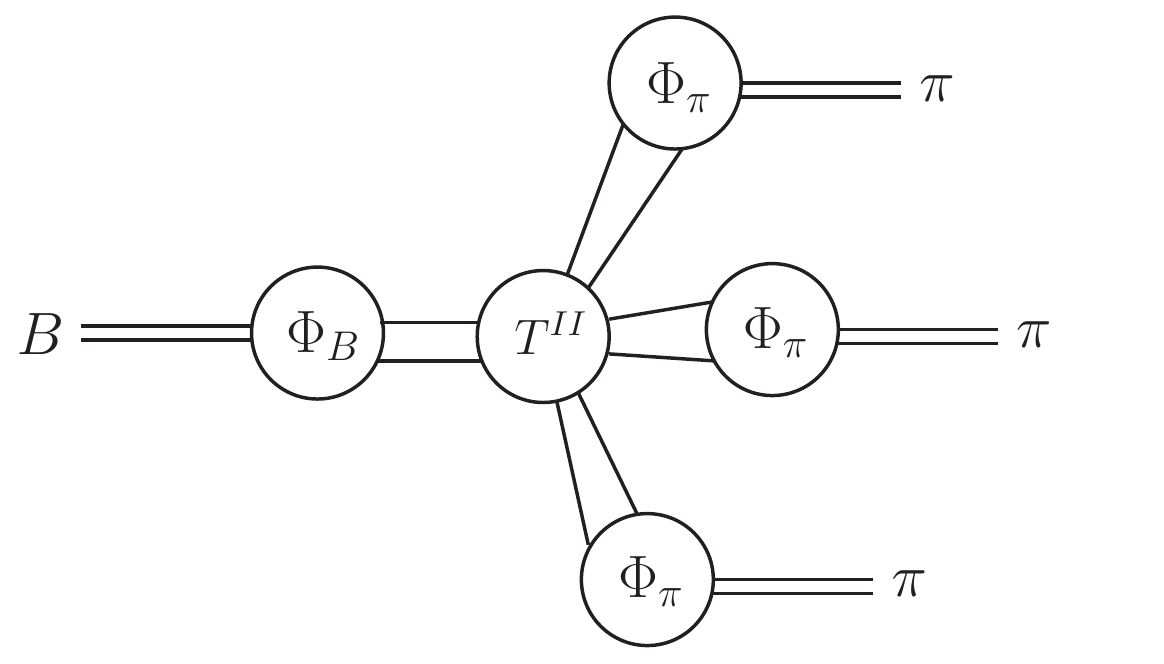}
\end{center}
\caption{Factorization formula in the center (region I) \label{fig:FactCenter}}
\end{figure}

\begin{figure}
\begin{center}
\rput(1.5,-0.7){(a)}
\rput(5.8,-0.7){(b)}
\rput(10.2,-0.7){(c)}
\rput(14.5,-0.7){(d)}
\includegraphics[width=3cm,height=3cm]{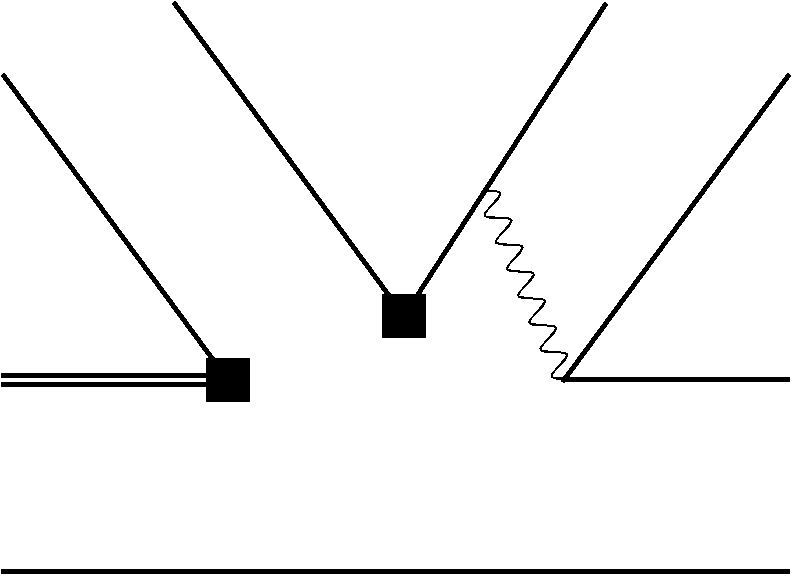}\hspace{12mm}
\includegraphics[width=3cm,height=3cm]{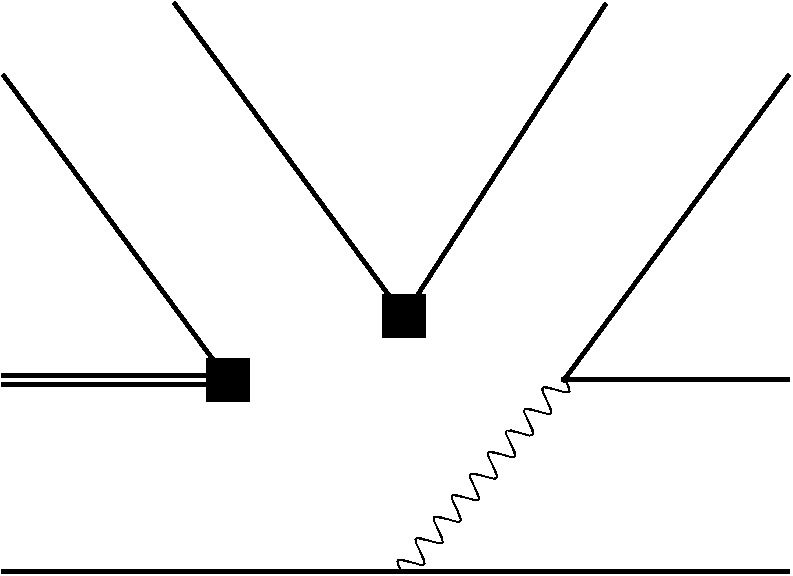}\hspace{12mm}
\includegraphics[width=3cm,height=3cm]{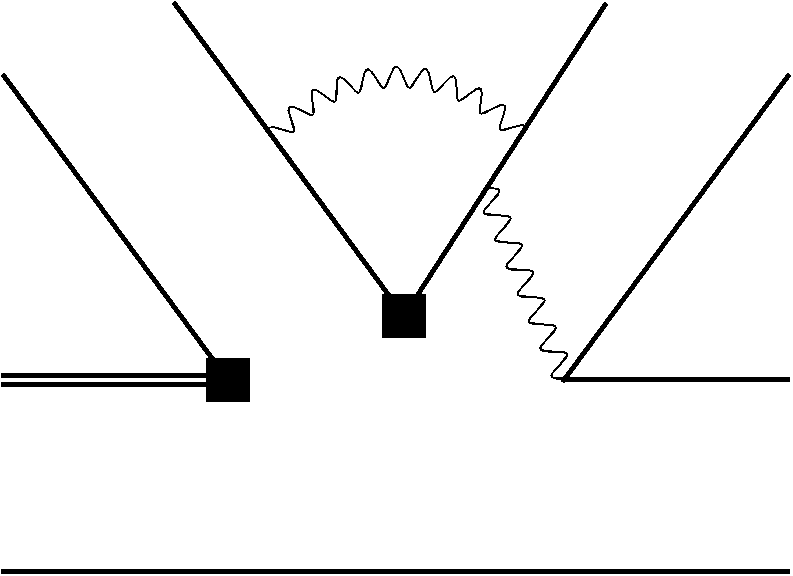}\hspace{12mm}
\includegraphics[width=3cm,height=3cm]{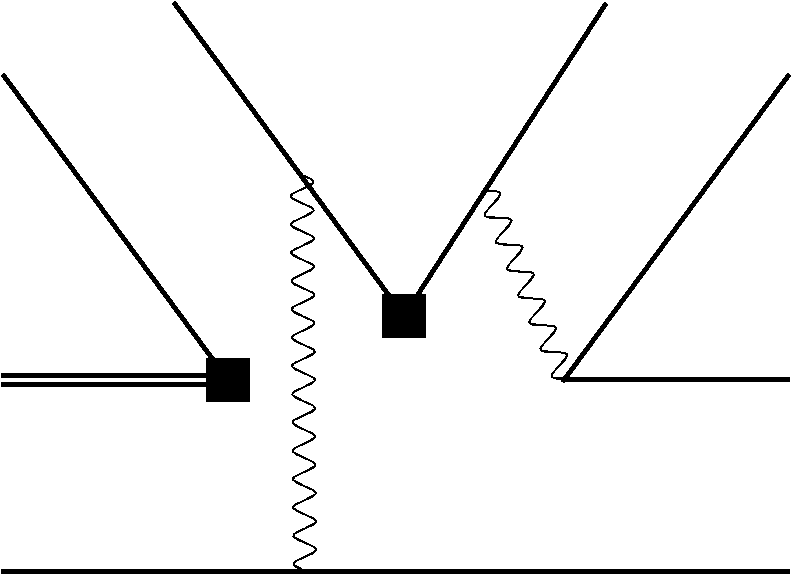}
\end{center}
\vspace{7mm}
\caption{Sample contributions to the hard kernels $T_i^I$ and $T_i^{II}$ in the factorization formula for the central region.
The leading $\alpha_s$ contributions are given by diagrams such as (a) and (b),
while (c) and (d) are next-to-leading in $\alpha_s$. See the text for details.
}
\label{fig:DiagsCenter}
\end{figure}

Our starting point is the heavy-quark limit, where we assume that $m_b / \sqrt{3} \gg \Lambda_{\rm QCD}$. In the central region of the Dalitz plot (region I) we have all invariant masses of the order $m_b / \sqrt{3}$
and hence we expect the factorization formula
\begin{equation} 
\langle \pi^+\pi^-\pi^+ | {\cal O}_i |B^+ \rangle_{s_{ij} \sim 1/3} = T_i^I \otimes F^{B\to \pi} \otimes \Phi_{\pi} \otimes \Phi_{\pi}
+ T_i^{II} \otimes \Phi_B \otimes \Phi_{\pi} \otimes \Phi_{\pi} \otimes \Phi_{\pi} \ ,
\label{eq:FactCenter}
\end{equation} 
where ${\cal O}_i $ is a four quark operator in the effective weak Hamiltonian. This factorization formula is illustrated in Fig.~\ref{fig:FactCenter}. 
The hard kernels can be computed perturbatively in QCD. Some typical diagrammatic contributions are shown in Fig.~\ref{fig:DiagsCenter}. We will consider here only the leading $\alpha_s$ corrections,
and neglect next-to-leading $\alpha_s^2$ contributions such as (c) and (d) in Fig.~\ref{fig:DiagsCenter}. While the study of $\alpha_s^2$ corrections is beyond the scope of this analysis,
we expect these to be about $\sim 10\%$ relative to the leading color-allowed amplitude,
similar to the case of $B\to\pi\pi$ (see e.g. Ref.~\cite{1501.07374}). The diagram (b), where the gluon is ejected from the spectator, requires the spectator quark in the $B$ meson to have a large
virtuality of order $m_b$, which is either suppressed in the heavy-quark limit or requires an additional hard interaction.
All in all, we do not include the second term in Eq.~(\ref{eq:FactCenter}) in our analysis, nor radiative corrections to $T_i^I$.
To this order, the convolutions of the hard kernel $T_i^I$ with the $B \to \pi$ form factor and the pion light-cone distribution can be computed without encountering end-point singularities.
While this would be a trivial statement in the case of two-body decays, we stress that here the kernels $T_i^I(u,v)$ already depend on the momentum fraction of the quarks at the leading
order, making the convolutions non-trivial.

The differential decay rate $d^2 \Gamma / (ds_{++} \, ds_{+-})$  computed in this way shows some interesting features.
First of all, moving from the central point $s_{++}=s_{+-}=1/3$ (region I) toward the edge $s_{++}\sim 0$ (region IIa),
we find that the rate remains regular, that is, it approaches a finite limit as $s_{++} \to 0$.
This can be seen explicitly in the calculation, with no propagator becoming soft as $s_{++} \to 0$.
More precisely, moving away from the center along the line $s_{+-}=(1-s_{++})/2$, we find
\begin{equation}
\frac{d \Gamma}{ds_{++} \, ds_{+-} } \Bigg|_{
\begin{minipage}{15mm}
\scriptsize
$s_{++}\to 0$\\[-1mm]
$s_{+-}\sim 1/2$\\[1mm]
\end{minipage}} \sim
\ \Gamma_0\ f_+(m_B^2/2)^2
\end{equation} 
up to a coefficient of order one, with
\begin{equation}
\Gamma_0 = \frac{G_F^2 \alpha_s^2 (m_b)  f_\pi^4 m_B |V_{ub}V_{ud}^*|^2 }{32\pi}\ .
\end{equation}
Here $f_+(q^2)$ denotes the vector $B\to\pi$ form factor, defined as:
\begin{equation}
q_\mu\,\av{\pi(p-q)| \bar b \gamma^\mu q|B(p)} = (m_B^2-m_\pi^2) f_0(q^2) \simeq (m_B^2-q^2)  f_+(q^2) \ ,
\end{equation}
where in the last term we have employed the large-recoil-energy relation \cite{0008255}.

\begin{figure}
\begin{center}
\includegraphics[width=8cm]{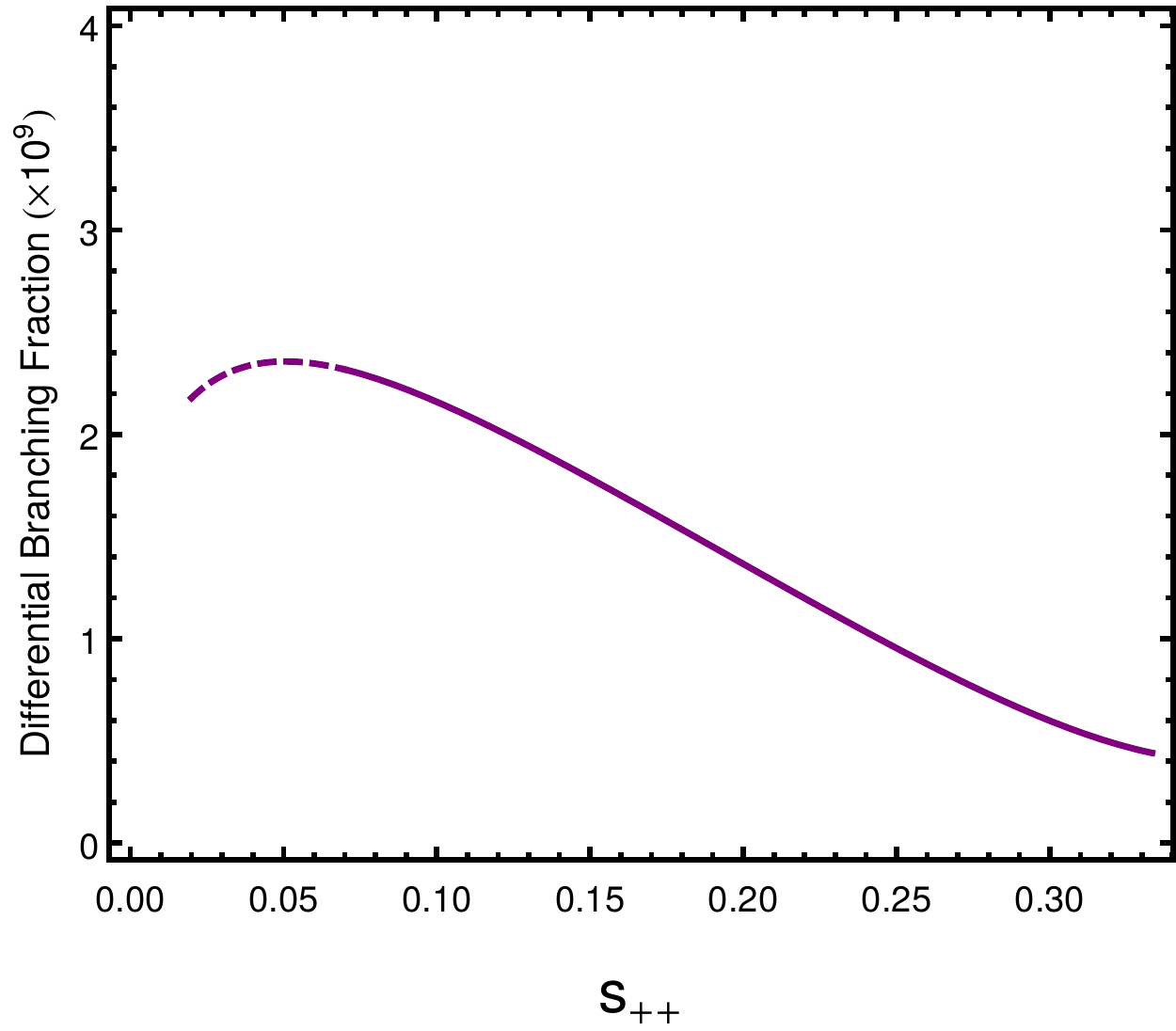}
\hspace{5mm}
\includegraphics[width=8cm]{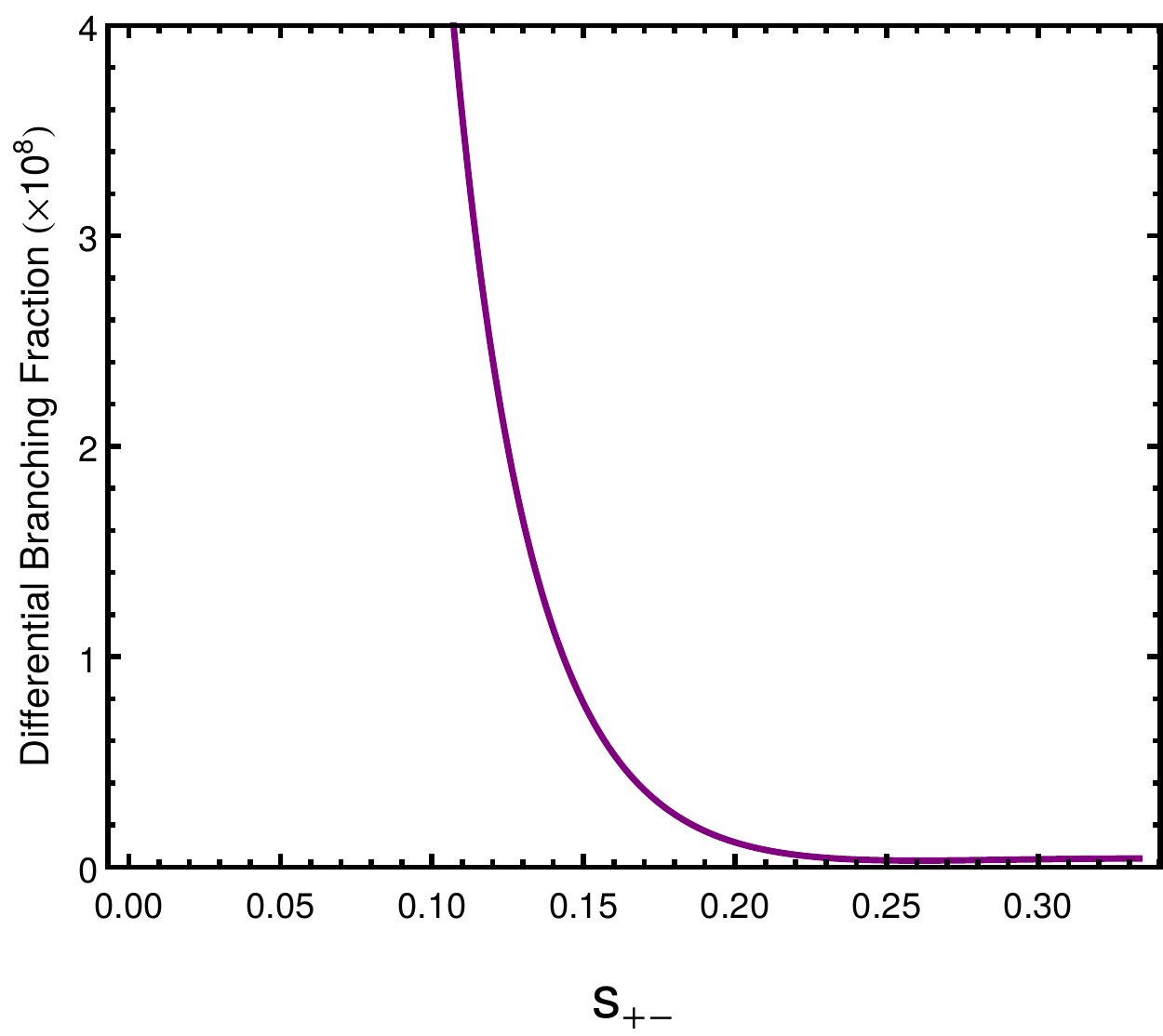}
\end{center}
\caption{Differential decay rate when extrapolated from the center region of the Dalitz plot towards the collinear edges.
The extrapolation to small $s_{++}$ remains regular, while the limit to small $s_{+-}$ diverges. See text for details.}
\label{fig:extrapolations}
\end{figure}

In Fig.~\ref{fig:extrapolations} (left panel) we show the exact dependence of the rate as a function of $s_{++}$,
along this direction in the Dalitz plane. This regular behavior does not depend on how we approach the $s_{++}=0$ edge.

The situation is very different if we consider the behavior of $d^2 \Gamma / (ds_{++} \, ds_{+-})$ as $s_{+-}$ gets small (towards region IIb). We consider now the direction along the line $s_{++}=(1-s_{+-})/2$. In this region the rate behaves as,
\begin{equation}
\frac{d \Gamma}{ds_{++} \, ds_{+-} } \Bigg|_{
\begin{minipage}{15mm}
\scriptsize
$s_{+-}\to 0$\\[-1mm]
$s_{++}\sim 1/2$\\[1mm]
\end{minipage}} \sim
\frac{1}{s_{+-}^2} \Gamma_0\ f_+(m_B^2/2)^2\ + \text{ regular terms as }  s_{+-} \to 0
\end{equation} 
rendering the rate non-integrable. This behavior is expected, as the edge of the Dalitz plot with small $s_{+-}$ is 
determined by hadronic resonances, dominantly the $\rho$ resonance. In this region the three-body decay effectively becomes a
quasi two-body  decay, and the methods to describe collinear and soft parts of the Dalitz plot have to be modified.

\section{The collinear regions of the Dalitz plot} 
\label{sec:edge}
The case where one invariant mass is small is kinematically very similar to a two-body decay. We expect then a
similar factorization theorem, with the difference that one of the particles in the two-body case is substituted by a pair of particles with small invariant mass, which must be described collectively.
In order to describe the soft and collinear regions of the Dalitz plot, we therefore need to introduce additional non-perturbative quantities:  the two-pion light-cone distribution (2$\pi$LCD) amplitude
and the $B \to \pi \pi$ from factor.   

\bigskip

To leading twist, the 2$\pi$LCD for a  2-pion system $(\pi^+ \pi^-)$ is formally given by the matrix element
\cite{Grozin:1983tt,9805380,9809483,0307382}
\begin{equation} 
S_{\alpha\beta}^q(z,k_1,k_2) = \frac{k_{12}^+}{4\pi} \int d x^- e^{-i z (k_{12}^+ x^-)/2}
\langle \pi^+(k_1)\pi^-(k_2)|\bar q_{\beta} (x) [x,0] q_{\alpha}(0)|0 \rangle_{x^+=x_\bot=0} 
\end{equation}
where $\{\alpha,\beta\}$ are Dirac indices, $q=u,d$, and $[x,0]$ is a Wilson line.
We take $k_{12}=k_1+k_2$ and define two light-like vectors $n_\pm^\mu = (1,0,0,\pm 1)$ such that
\begin{equation}
k_{12}^\mu = \frac{k_{12}^+}2 n_+^\mu + \frac{k_{12}^-}2 n_-^\mu \quad \text{and} \quad  x^\mu = \frac{x^+}2 n_+^\mu + \frac{x^-}2 n_-^\mu + x_\bot^\mu\ ,
\end{equation}
and such that when $k_{12}^2\to 0$, then $k_{12}^-\to 0$. The variable $z$ is the fraction of the momentum
$k_{12}$ carried by the quark $q$. 

The Lorentz decomposition of the matrices $S_{\alpha\beta}^q$ consistent with parity invariance, keeping only terms that contribute at twist-2, is
given by\footnote{This definition of $\Phi_\|$ agrees with reference \cite{9809483} (up to isospin decomposition, see later).
However the definition for $\Phi_\bot$ might differ from that in \cite{9809483} by an overall
factor, which we do not address here because to the order considered $\Phi_\bot$ will not appear in the amplitude.}
\begin{equation} 
S_{\alpha\beta}^q = \frac14 \Phi^q_\|(z,\zeta,k_{12}^2)\,\sla{k}_{12}
+ \Phi^q_\bot(z,\zeta,k_{12}^2)\,\sigma_{\mu\nu}k_1^\mu k_2^\nu\ ,
\end{equation} 
which defines the vector ($\Phi_\|$) and tensor ($\Phi_\bot$) 2$\pi$LCDs.
The variable $\zeta=k_1^+/k_{12}^+$ is the light-cone momentum fraction of $\pi^+$.
In terms of invariants, we have
\begin{equation}
k_{12}^2 = m_B^2\,s_{12}\ ,\qquad \zeta = \frac{s_{13}}{1-s_{12}}\ .
\label{zeta}
\end{equation} 

Isosinglet ($\Phi^{\bf 0}\equiv \frac12[\Phi^{\bf u} + \Phi^{\bf d}]$) and isovector
($\Phi^{\bf 1}\equiv \frac12[\Phi^{\bf u} - \Phi^{\bf d}]$) 2$\pi$LCDs have been discussed in the
literature (e.g. Refs~\cite{9809483,0307382}). The vector $I=1$ $2\pi$LCDs are normalized as \cite{9809483}:
\begin{eqnarray} 
\int dz \,\Phi^\text{\bf 1}_\|(z,\zeta,s) &=& (2\zeta - 1) F_\pi(s) \label{F(s)}\ ,
\end{eqnarray}
where $F_\pi(s)$ denotes the pion vector form factor. $C$-parity and isospin invariance imply that the corresponding integral
is zero for the isosinglet component \cite{9809483}.
At the leading order, the hard kernel $T(z)$ with which $\Phi(z,\zeta,s)$ is convoluted in the $B\to \pi\pi\pi$ amplitude 
does not depend on the momentum fraction $z$, so the amplitude depends only on the local form factor $F_\pi(s)$,
just as the leading contribution in $B\to \pi\pi$ depends only on $f_\pi$. In addition, at the leading order the tensor
distribution $\Phi_T$ does not contribute.

\begin{figure}
\includegraphics[width=8.2cm,height=8cm]{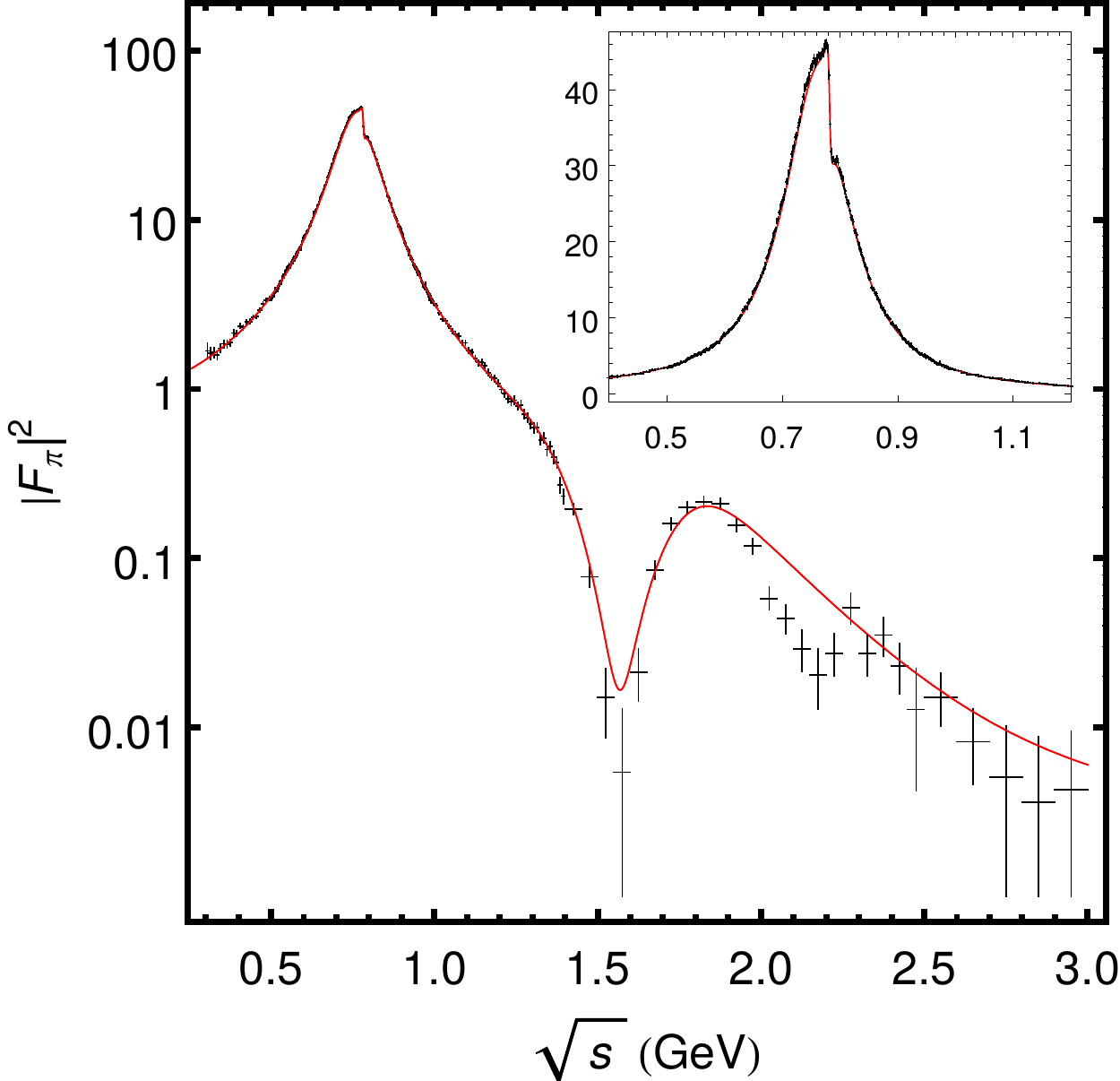}\hspace{5mm}
\includegraphics[width=8.2cm,height=8cm]{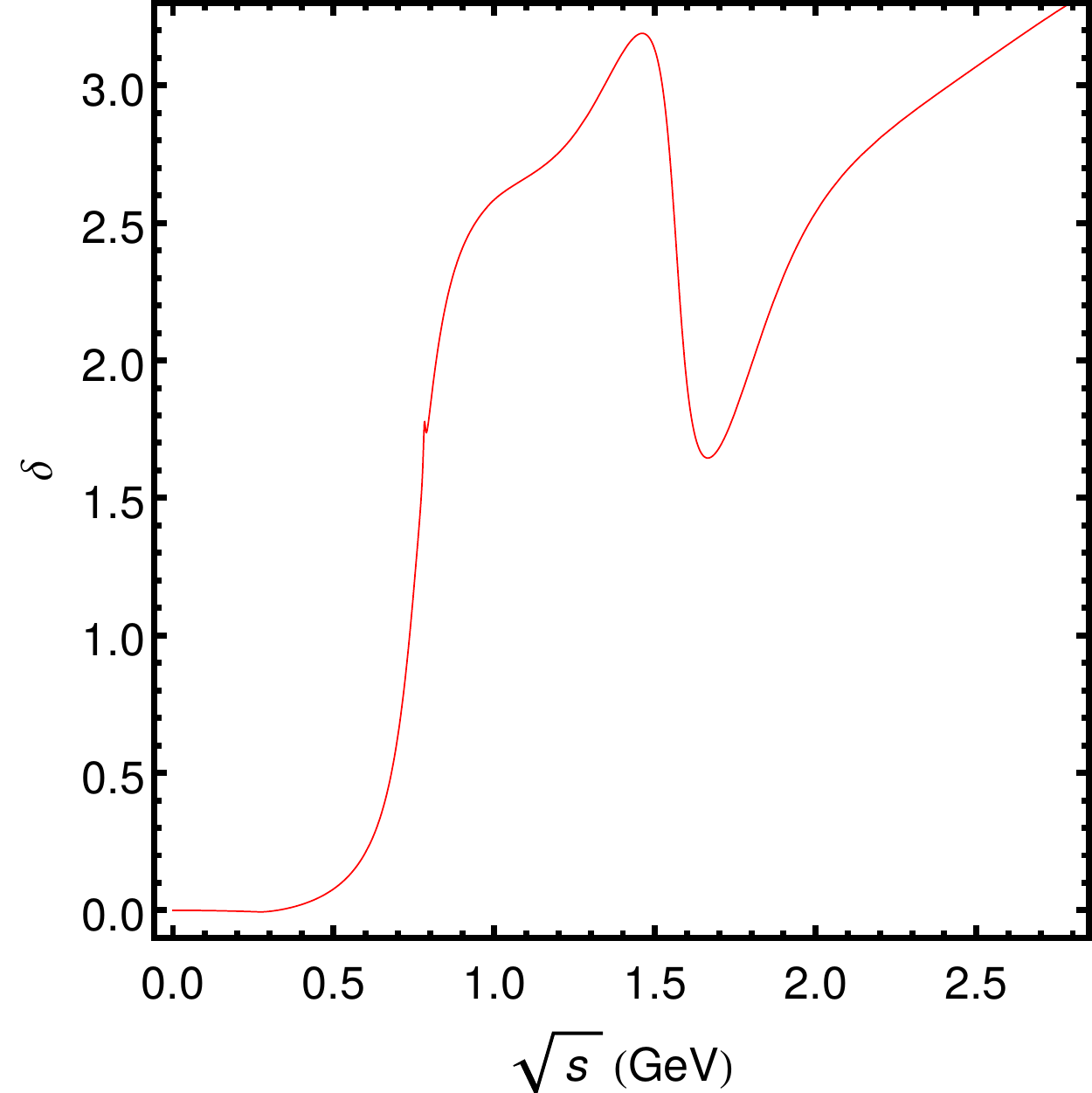}
\caption{Pion form factor $F_\pi(s)=|F_\pi| e^{i\delta}$ in the time-like region \cite{1205.2228,1203.3955}.}
\label{pionFF}
\end{figure}

The vector form factor $F_\pi(s)$ in the time-like region ($s>0$) can be obtained from measurements of the process $e^+e^- \to \pi^+\pi^-(\gamma)$~\cite{1205.2228}\footnote{
For simplicity we use only the latest Babar data, but see also Refs.~\cite{0308008,0610021,0605013,0407048,0809.3950,1006.5313,1105.4975}.
} -- see Fig.~\ref{pionFF}.
We employ here the fit parametrization of Ref.~\cite{1203.3955}, which is consistent with general principles of QCD at low energies,
and covers the energy range of interest, including the relevant resonances in that range.
The particular choice of parametrization is not very important for the absolute value $|F_\pi(s)|$, where a good fit to the data is enough (see Fig.~\ref{pionFF}), but
it is important for the phase, where data is not so precise. A thorough analysis of the phase of $F_\pi(s)$ and its impact in $B\to\pi\pi\pi$ is beyond the scope of this
paper, but it becomes a crucial issue as soon as one attempts to describe CP asymmetries. We leave this for future work.

\bigskip

The second nonperturbative input is the
$B \to \pi \pi $ form factor, which  has been discussed already in the context of $B\to \pi\pi\ell\nu$ decays in
Ref.~\cite{1310.6660}. We consider the generic form factor:
\begin{equation} 
F_{\alpha\beta}(k_1,k_2,k_3) \equiv \langle \pi^+(k_1)\pi^-(k_2)|\,\bar b_\beta\, u_\alpha|B^+(p) \rangle \,  ,
\end{equation} 
where $\alpha,\beta$ are Dirac indices. The most general Lorentz decomposition consistent with parity invariance is given
in terms of four independent form factors:
\begin{equation} 
F = F_t \frac1{4\sqrt{k_3^2}} \sla k_3\gamma_5
+ F_2\, \sla k_{(0)}\gamma_5
+ F_3\, \sla {\bar k}_{(\|)}\gamma_5
+ F_4\, \epsilon_{\alpha\beta\gamma\mu} k_1^\alpha k_2^\beta k_3^\gamma\gamma^\mu
+\frac{\sqrt{k_3^2}}{4(m_b+m_u)}  F_t \gamma_5 \ ,
\label{eq:F}
\end{equation} 
with $k_{(0)}, k_{(\|)}$ two orthogonal space-like vectors. Due to the structure of the leading order contributions, the time-like form
factor $F_t(\zeta,s_{12})$ will be the only one relevant here. This definition of $F_t$ coincides with Ref.~\cite{1310.6660}.

In order to be able to make a quantitative prediction, we can relate the different $B\to\pi\pi$ form factors to the $2\pi$LCDs
via a light-cone sum rule \cite{HK}. For the time-like form factor $F_t$ we have\footnote{
This is a tentative expression where we have ignored a possible contribution from the distribution $\Phi_\bot$.
We use this formula for illustrative purposes. The final form of this expression will be presented in Ref.~\cite{HK}.}:
\begin{equation}  \label{QCDSR} 
F_t(\zeta,s_{12}) = \frac{m_b^2}{\sqrt{2}\hat f_B \sqrt{k_3^2}} \int_{u_0}^1 \frac{dz}{z}\,
\exp \left[\frac{(1+s_{12}\bar z)m_B^2}{M^2} - \frac{m_b^2}{z M^2}\right]\,
\Phi_\|(z,\zeta,s_{12})\ ,
\end{equation} 
where $\hat f_B$ is the static $B$-meson decay constant extracted from a corresponding sum-rule, which is correlated to
the Borel parameter $M$ and to the threshold parameter $u_0$. These three parameters must be determined simultaneously
with the condition that the physical decay constant and form factor are independent of $M$ and $u_0$. While we do not
attempt to perform a full error analysis here, we note that the values $\hat f_B \simeq 0.316$, $u_0\simeq 0.6$ and
$M^2 \simeq 10$~GeV$^2$ satisfy this correlation approximately.
In the asymptotic limit, given by $\Phi_\|^{\bf 1} = 6 z(1-z)(2\zeta - 1) F_\pi(s)$ \cite{0307382},
and setting $\sqrt{k_3^2}=m_\pi$, we have
\begin{equation} 
F_t^{\bf 1}(\zeta,s_{12}) = \frac{3\sqrt{2} m_b^2 (2\zeta - 1) F_\pi(s_{12})}{\hat f_B m_\pi} \int_{u_0}^1 dz\ \bar z\,
\exp \left[\frac{(1+s_{12}\bar z)m_B^2}{M^2} - \frac{m_b^2}{z M^2}\right]\ .
\end{equation} 

\begin{figure}
\begin{center}
\includegraphics[width=5cm,height=3.3cm]{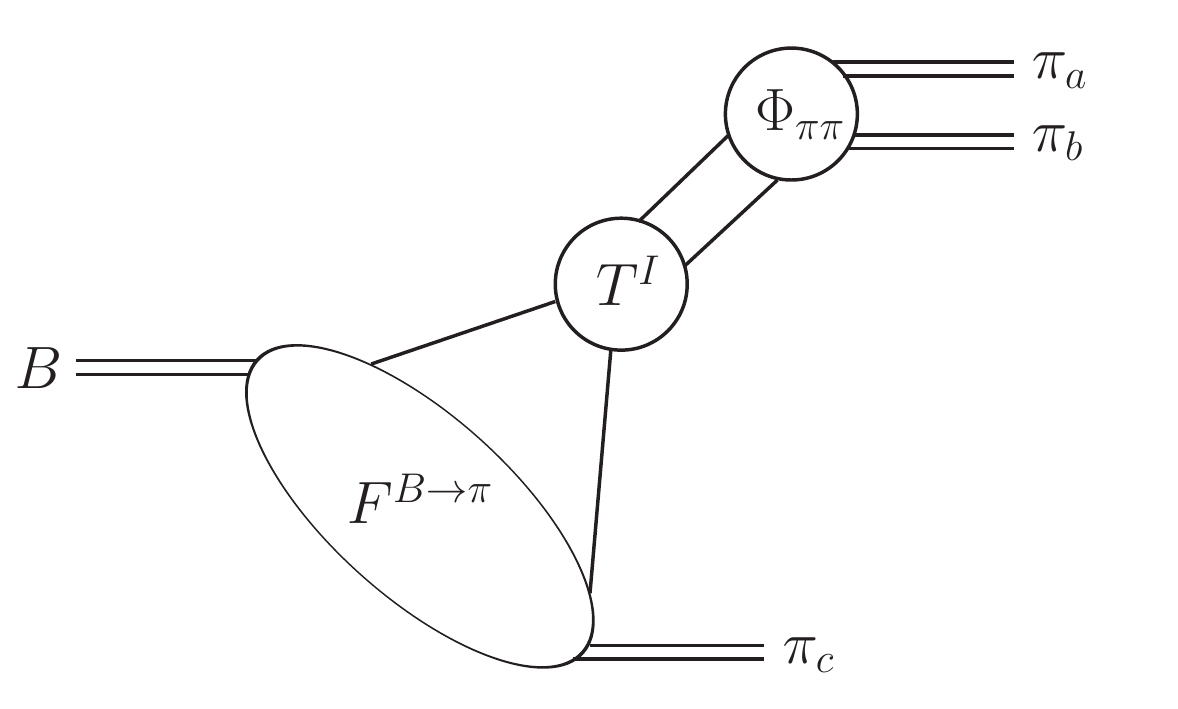} 
\hspace{-5mm}
\raisebox{12mm}{+}
\hspace{5mm}
\includegraphics[width=5cm,height=3.3cm]{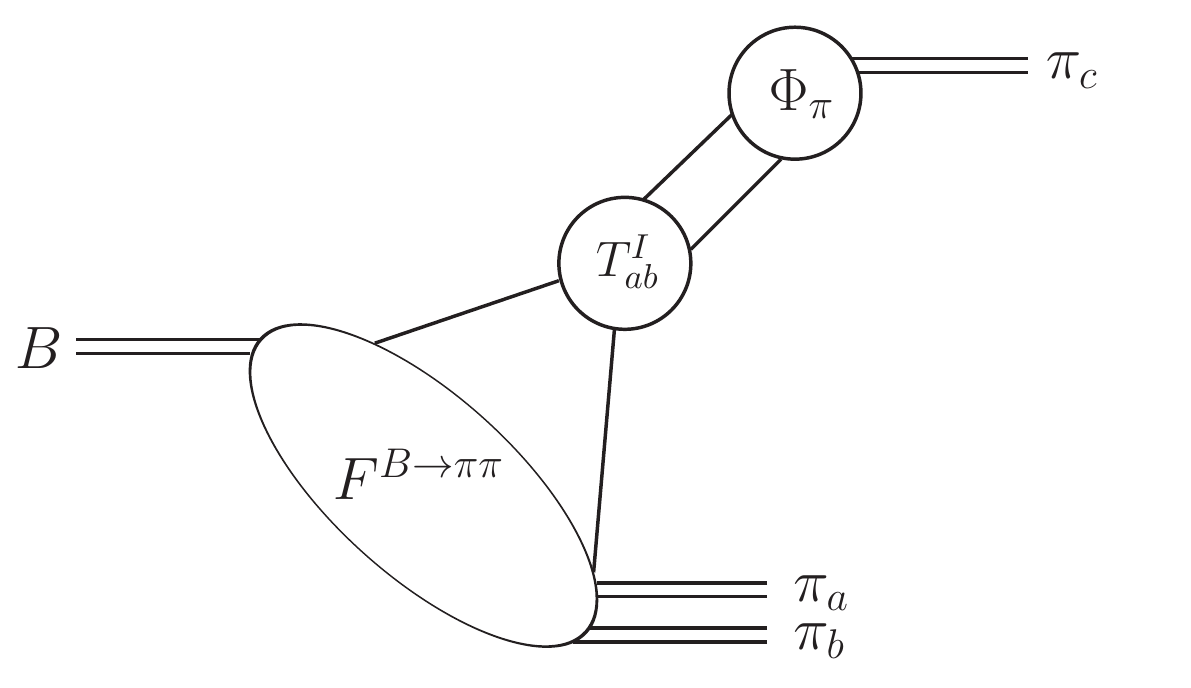}
\hspace{-5mm}
\raisebox{12mm}{+}
\hspace{5mm}
\includegraphics[width=5cm,height=3.0cm]{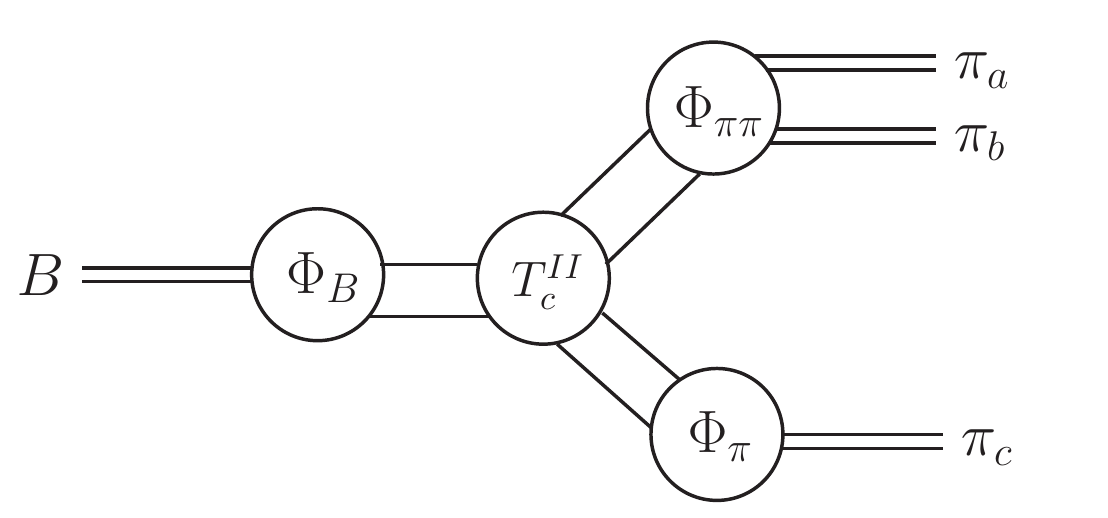}
\end{center}
\caption{Factorization formula for the edges of the Dalitz plot (region II). \label{fig:s23large}}
\label{fig:s23small}
\end{figure}

With this we have all ingredients for the factorization formula valid in the collinear regions of the Dalitz plot. 
The modified QCD factorization formula reads, in terms of the new non-perturbative quantities:  
\begin{eqnarray} 
\av{\pi^a\pi^b\pi^c|\op_i|B}_{s_{ab} \ll 1} &=& T_c^I \otimes F^{B\to \pi^c} \otimes \Phi_{\pi^a\pi^b}
+ T_{ab}^I \otimes F^{B\to \pi^a\pi^b} \otimes \Phi_{\pi^c}\nn\\[2mm]
&+& T^{II} \otimes \Phi_B \otimes \Phi_{\pi^c} \otimes \Phi_{\pi^a\pi^b}\ .
\label{s23small}
\end{eqnarray}
This formula is illustrated in Fig.~\ref{fig:s23small} and yields now the description of the Dalitz plot in the kinematic regions IIa and IIb in  Fig.~\ref{Dalitz}.

Using the QCD sum rule relation (\ref{QCDSR}) for the $B\to \pi\pi$ form factor, and writing the 2$\pi$LCD
in terms of the pion form factor $F_\pi$, we may now write down the amplitude in the region of small $s_{+-}$, at leading order and leading twist. We find
\footnote{Here $F_t$ corresponds to the combination $F_t^{\rm 0} + F_t^{\rm 1}$. The convolution with $\Phi_\|^{\bf 0}$ in the sum-rule for $F_t^{\rm 0}$ is in general not
zero because the integrand is not even in $z$. Since the isosinglet distribution is mostly unknown, even in the asymptotic limit, we will nevertheless disregard this term altogether in the numerical
analysis, keeping in mind that this issue requires further investigation.}:
\begin{equation}
\A|_{s_{+-} \ll 1} = \frac{G_F}{\sqrt{2}} \big[  f_\pi m_\pi (a_1-a_4) \cdot F_t(\zeta,s_{+-})  + m_B^2 (a_2+a_4) (2\zeta -1) \cdot f_0(s_{+-}) \cdot F_\pi(s_{+-}) \big]\ ,
\label{eq:Aedge}
\end{equation}
where the parameters $a_i$ are combinations of Wilson coefficients\footnote{More specifically, we have $a_{1,2} = V_{ub}^* V_{ud} (C_{1,2}+C_{2,1}/N_c)$
and $a_{3,4} = V_{tb}^* V_{td} (C_{3,4}+C_{4,3}/N_c)$, with the Wilson coefficients $C_i$ defined as in Ref.~\cite{0104110}.} (see e.g. Ref.~\cite{0104110}).
To this order all the convolution integrals are trivial. Again, we neglect $\alpha_s$ corrections and hard-scattering with the spectator quark for simplicity.
Hard kernels are known already at NNLO from studies of two-body decays \cite{0512351,0608291,0610322,0705.3127,0709.3214,0911.3655,1410.2804}, but the convolutions
with two-pion distributions still need to be worked out. In particular new distributions appear (e.g. $\Phi_\bot$) that do not contribute at the leading order.
This is beyond the scope of this work. The conclusions derived here at this order of approximation should nevertheless remain valid.

A qualitative difference of three-body decays in this kinematic regime with respect to two-body decays is that the nonperturbative input is much richer in terms of QCD effects.
In particular, $F_\pi$ contains resonance and rescattering contributions, including an imaginary part from non-perturbative dynamics, in contrast to two-body decays where strong phases are, at the
leading power, of perturbative origin. This has implications both for quasi-two-body decays and for CP asymmetries.
Most of the information on $F_\pi(s)$ can be obtained from data (see Fig.~\ref{pionFF}), allowing for
a data-driven model-independent interpretation of three-body Dalitz plots, at least within the accuracy of factorization theorems.

\begin{figure}
\begin{center}
\includegraphics[width=8.2cm]{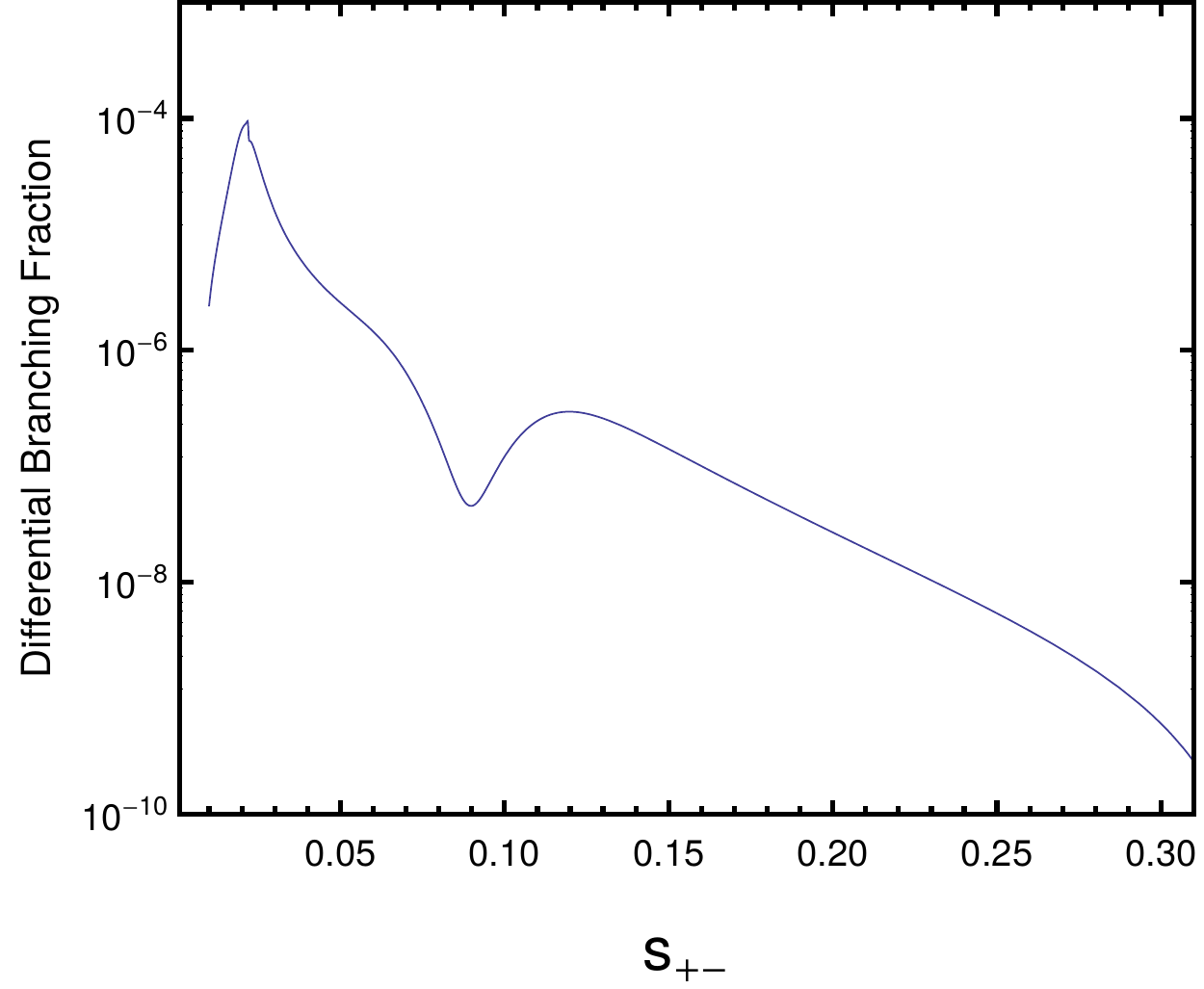}
\hspace{5mm}
\includegraphics[width=8cm]{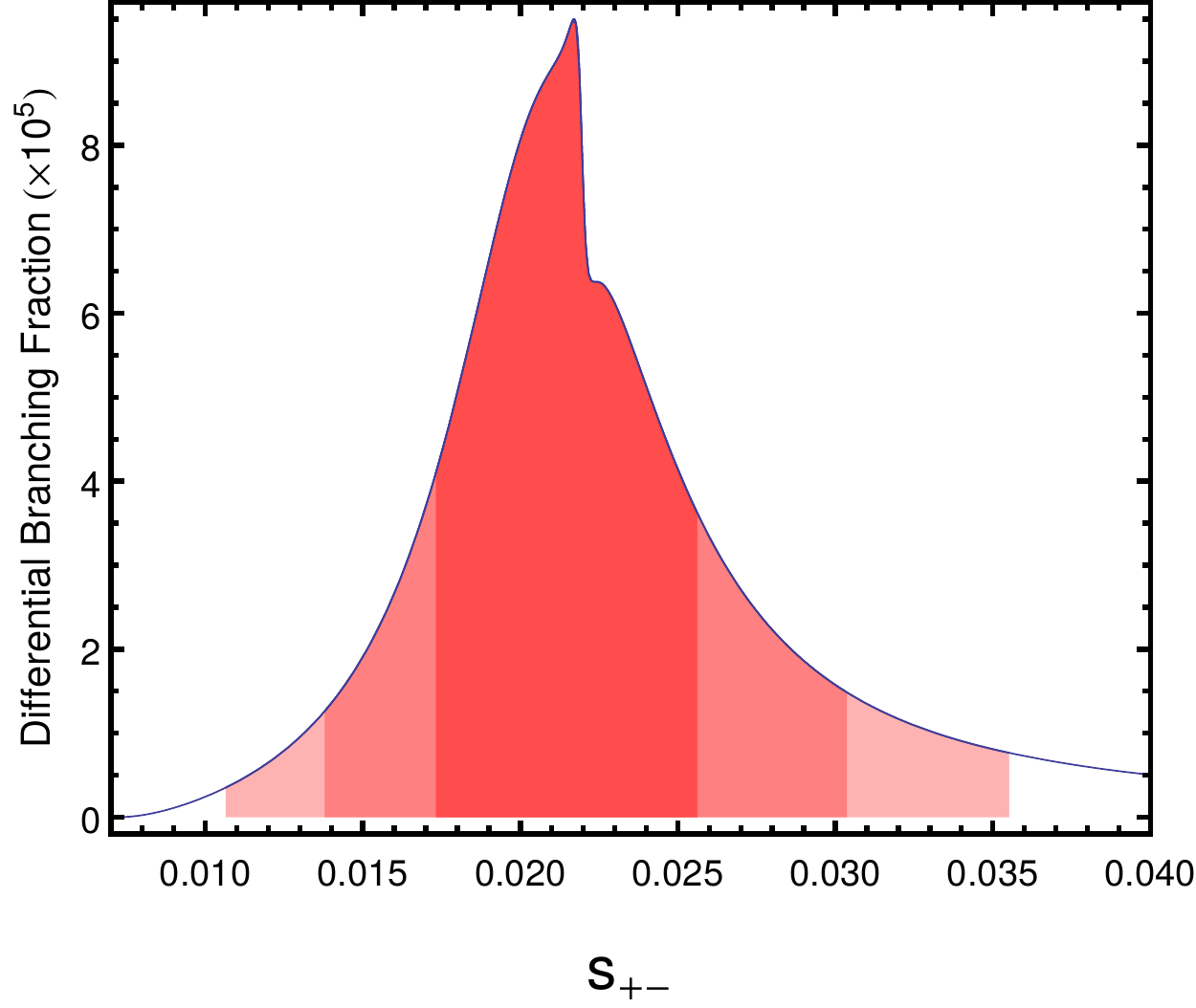}
\end{center}
\caption{Differential decay rate obtained from the description in terms of two-pion distributions for small $s_{+-}$.
Left: extrapolation to $s_{+-}\sim 1/3$, with the $\rho''(1700)$ apparent, and the $\rho-\omega-\rho'$ peak, in logarithmic scale. Right: Zoom to resonant contribution from the
$\rho(770)$ and $\omega(782)$. }
\label{fig:edge}
\end{figure}

As a simple application of this result, we estimate the branching fraction $BR(B^+\to \rho \pi^+)$ by integrating the
differential decay rate in a neighborhood of the $\rho$ resonance:
\begin{equation}
\widehat{BR} (B^+\to \rho \pi^+) = \int_{0}^{1} ds_{++} \int_{s_\rho^-}^{s_\rho^+} ds_{+-}\ 
\frac{\tau_B\,d\Gamma}{ds_{++}ds_{+-}}
= \int_{0}^{1} ds_{++} \int_{s_\rho^-}^{s_\rho^+} ds_{+-}\ 
\frac{\tau_B\,m_B |\A|^2}{32(2\pi)^3}\ ,
\end{equation}
where $s_\rho^\pm = (m_\rho \pm \delta)^2/m_B^2$, and we will take $\delta = n \Gamma_\rho$, with $n$ specifying the cuts in units of the $\rho$-meson width.
We find:
\begin{eqnarray}
\widehat{BR}(B^+\to \rho \pi^+) \simeq 2.4\cdot 10^{-6} && \text{for }\ n=1\\
\widehat{BR}(B^+\to \rho \pi^+) \simeq 3.0\cdot 10^{-6} && \text{for }\ n=2\\
\widehat{BR}(B^+\to \rho \pi^+) \simeq 3.2\cdot 10^{-6} && \text{for }\ n=3\\
\widehat{BR}(B^+\to \rho \pi^+) \simeq 3.3\cdot 10^{-6} && \text{for }\ n=4
\end{eqnarray}
We note that extending the cuts beyond $m_\rho \pm 4\Gamma_\rho$ does not modify the result very much, as the resonant $\rho$ contribution dominates the full decay rate.
Comparing these numbers to the experimental value \cite{PDG},
\begin{equation}
BR(B^+\to \rho \pi^+)^\text{exp} = (8.3\pm 1.2)\cdot 10^{-6}
\end{equation}
we see that the result is in the right ballpark. However, $\widehat{BR}$ is an object different from the $B\to \rho\pi$ branching fraction as given in \cite{PDG},
and can be measured experimentally in a direct and model-independent manner, without the need to extract the $\rho$ from the
full distribution. At this point we must emphasize that this is still a very crude estimate, and a more careful study would need to be performed
to really test the data.

\section{Discussion}
So far we have used two different factorization formulas for region I and region II. Region I has been described using the conventional QCD factorization
in terms of single pion states (which we will call QCDF$_\text{I}$ hereafter), while region II has been described in terms of hadronic input describing two-pion states
with small invariant mass (called QCDF$_\text{II}$ hereon). 
To get the full Dalitz distribution one needs to match the result from the central region with the one of the edges.
To this end, we assume that there is an intermediate region between the edge ($s_{+-}^\text{low}\equiv s \simeq 0$)
and the center ($s_{+-}^\text{low}\equiv s \simeq s_{+-}^\text{high} \simeq 1/3$) where both descriptions apply.
This region corresponds to $\Lambda_\text{QCD}^2/m_B^2 \ll s \ll 1/3$, and 
it certainly exists if $m_B$ is large enough. We will investigate below whether this happens for realistic $B$-meson masses.

In this intermediate region, one might use QCDF$_\text{II}$ (as in Section~\ref{sec:edge}) to write the amplitude in terms of two-pion states,
then take the perturbative limit for the $2\pi$LCDs and $B\to\pi\pi$ form factors, and finally compare the result with the
factorized QCDF$_\text{I}$ amplitude of Section~\ref{sec:MS}.
The idea is that, for  $s \gg \Lambda_\text{QCD}^2/m_B^2$, we have (schematically)\footnote{The factorization of 2$\pi$LCDs in
the perturbative limit has been studied in Ref.~\cite{9912364}. The factorization of $B\to\pi\pi$ form factors will be discussed
in \cite{PB-TF-DvD}.
The dots in Eq.~(\ref{FBpipiFact}) account for ``factorizable" contributions proportional to the $B$-meson light-cone distribution,
corresponding to neglected contributions in Section~\ref{sec:MS}.}:
\begin{eqnarray}
\Phi_{\pi\pi} &\to& f_\pi^2\, \int du\,dv\, T_\phi(u,v)\,\phi_\pi(u)\,\phi_\pi(v)\ ,\label{PhipipiFact}\\
F^{B\to\pi\pi} &\to& f_\pi\,F^{B\to\pi}(0) \int du\, T_F(u,v)\,\phi_\pi(u) + \cdots\ .\label{FBpipiFact}
\end{eqnarray}
Taking this limit for the leading power contribution in QCDF$_\text{II}$, one recovers fully some of the contributions obtained
using QCDF$_\text{I}$. 

In Table~\ref{tab:correspondence} we show the correspondence between the different contributions to the amplitude in this intermediate region,
either in QCDF$_\text{I}$ or QCDF$_\text{II}$.
The first column shows the contributions from two-pion distribution amplitudes. In QCDF$_\text{II}$ (lower diagram), this is a leading-power
contribution proportional to the $2\pi$LCD, $\Phi_{\pi\pi}$. As the invariant mass of the two pions in this intermediate region is also large,
the two pions can be factorized according to Eq.~(\ref{PhipipiFact}).
The production of two pions with large invariant mass requires a hard gluon, as shown in by the diagram at the top (corresponding to QCDF$_\text{I}$).
A similar argument goes through for the $B\to\pi\pi$ contribution, shown in the second column. The contribution in QCDF$_\text{I}$
(where the two pions are assumed to have large invariant mass) requires a hard gluon (top diagram),
and can be obtained from the contribution in QCDF$_\text{II}$ (bottom diagram) by factorization of $F^{B\to\pi\pi}$
according to Eq.~(\ref{FBpipiFact}). We have checked analytically that applying Eqs.~(\ref{PhipipiFact}),~(\ref{FBpipiFact})
to the amplitude in Eq.~(\ref{eq:Aedge}) we recover the corresponding results in Section~\ref{sec:MS}.

However, some contributions in QCDF$_\text{I}$ correspond
to contributions in QCDF$_\text{II}$ that are power suppressed, and do not arise from the perturbative limit of leading power
contributions in QCDF$_\text{II}$.
These are shown in the last four columns in Table~\ref{tab:correspondence}. Again, the contributions in QCDF$_\text{I}$ ($s\gg \Lambda_\text{QCD}/m_B$)
require a hard gluon. Columns 3 and 4 show the cases in which this gluon becomes collinear (in the $[\pi\pi]$ direction)
as $s\to 0$. They are termed ``non-factorizable'' since the gluon connects the two different collinear sectors. As $s\to 0$,
the quark propagator remains hard, which represents a power suppression with respect to the leading contributions. Columns 5
and 6 show the cases in which the gluon remains hard for all $s<1/3$. For $s\to 0$, these match onto 6-quark operators that are again
power-suppressed with respect to the leading contributions.
There is therefore a one-to-one diagrammatic correspondence between QCDF$_\text{I}$ and QCDF$_\text{II}$, but
this correspondence does not respect the power counting.

\begin{table}
\centering
\begin{tabular}{|c|c|c|c|c|c|c|}
\hline
&&&&&&\\
\rotatebox{90}{\hspace{-12mm}Center (QCDF$_\text{I}$)}
&
\begin{minipage}{2.2cm}
\centering
\begin{pspicture}(0.8,2.5)(0,0)
\includegraphics[width=1.45cm,height=1.8cm]{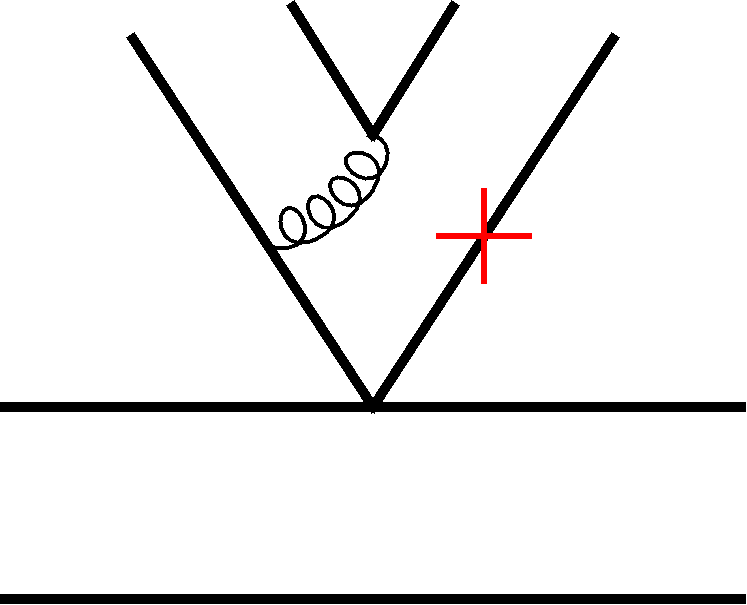}
\rput(-0.72,2.4){$\overbrace{\hspace{10mm}}^{\phi_{\pi\pi}}$}
\rput(0.5,0.27){$\Big\} {\scriptstyle F_{\pi}}$}
\end{pspicture}
\vspace{2mm}
\end{minipage}
&
\begin{minipage}{2.3cm}
\centering
\begin{pspicture}(0.8,2.5)(0,0)
\includegraphics[width=1.45cm,height=1.8cm]{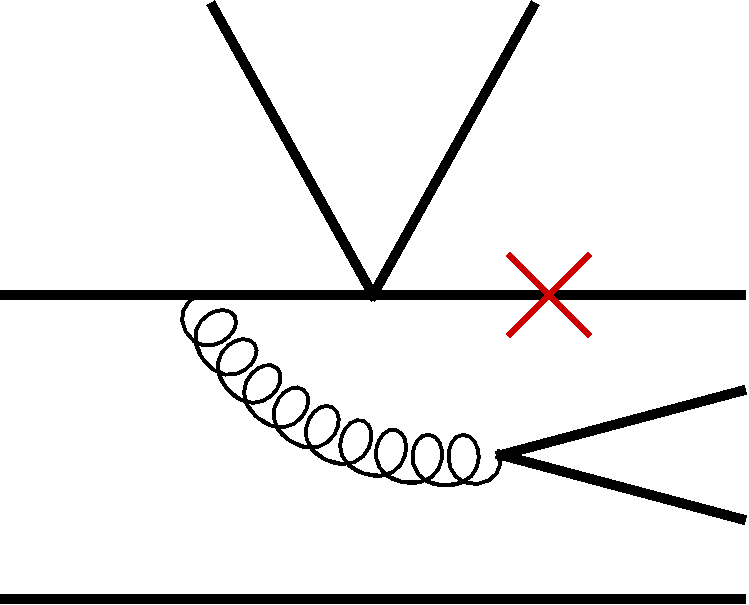}
\rput(-0.72,2.4){$\overbrace{\hspace{8mm}}^{\phi_{\pi}}$}
\rput(0.5,0.47){$\bigg\} {\scriptstyle F_{\pi\pi}}$}
\end{pspicture}
\vspace{2mm}
\end{minipage}
&
\begin{minipage}{2.3cm}
\centering
\begin{pspicture}(0.8,2.5)(0,0)
\includegraphics[width=1.45cm,height=1.8cm]{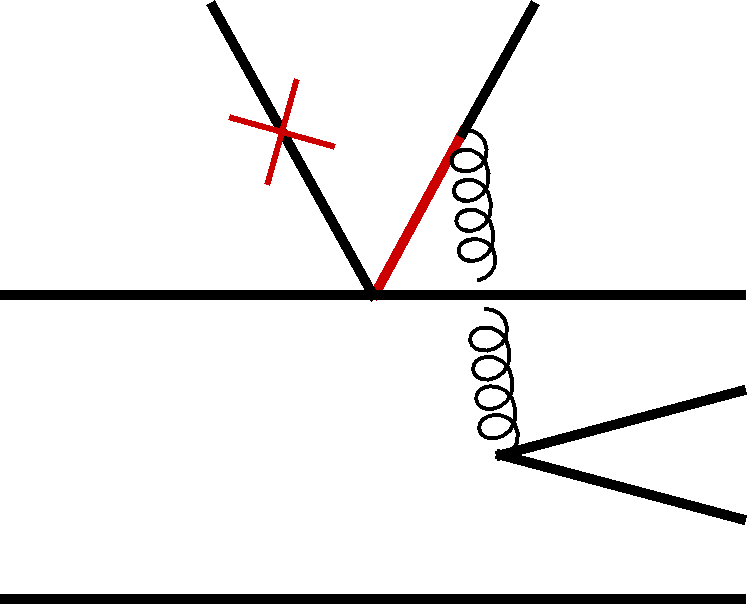}
\rput(-0.72,2.4){$\overbrace{\hspace{8mm}}^{\pi}$}
\rput(0.5,0.47){$\bigg\} {\scriptstyle \pi\pi}$}
\end{pspicture}
\vspace{2mm}
\end{minipage}
&
\begin{minipage}{2.2cm}
\centering
\begin{pspicture}(0.8,2.5)(0,0)
\includegraphics[width=1.45cm,height=1.8cm]{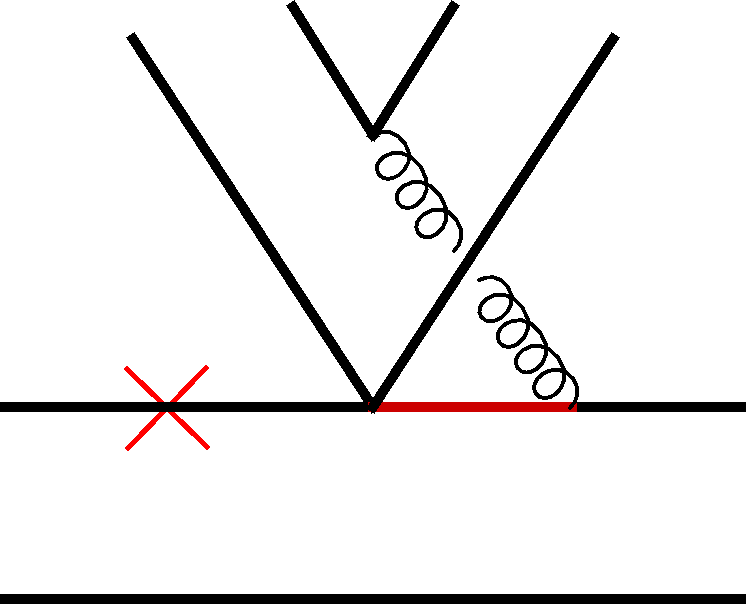}
\rput(-0.72,2.4){$\overbrace{\hspace{10mm}}^{\pi\pi}$}
\rput(0.5,0.27){$\Big\} {\scriptstyle \pi}$}
\end{pspicture}
\vspace{2mm}
\end{minipage}
&
\begin{minipage}{2.4cm}
\centering
\begin{pspicture}(0.8,2)(0,0)
\includegraphics[width=1.45cm,height=1.8cm]{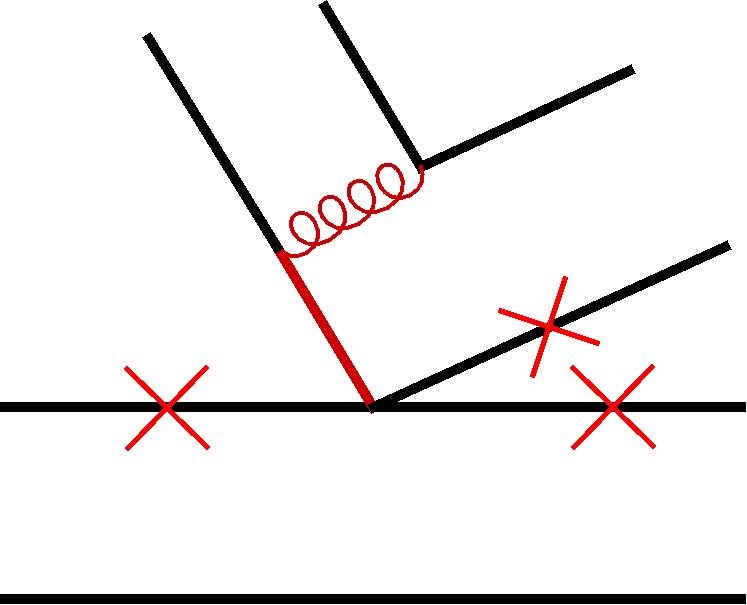}
\rput(-1,2.4){$\overbrace{\hspace{8mm}}^{\pi}$}
\rput(0.5,0.8){$\left.\begin{minipage}{1mm}\vspace{17mm}\end{minipage}\right\} {\scriptstyle \pi\pi}$}
\end{pspicture}
\vspace{2mm}
\end{minipage}
&
\begin{minipage}{2.2cm}
\centering
\begin{pspicture}(0.8,2.5)(0,0)
\includegraphics[width=1.45cm,height=1.8cm]{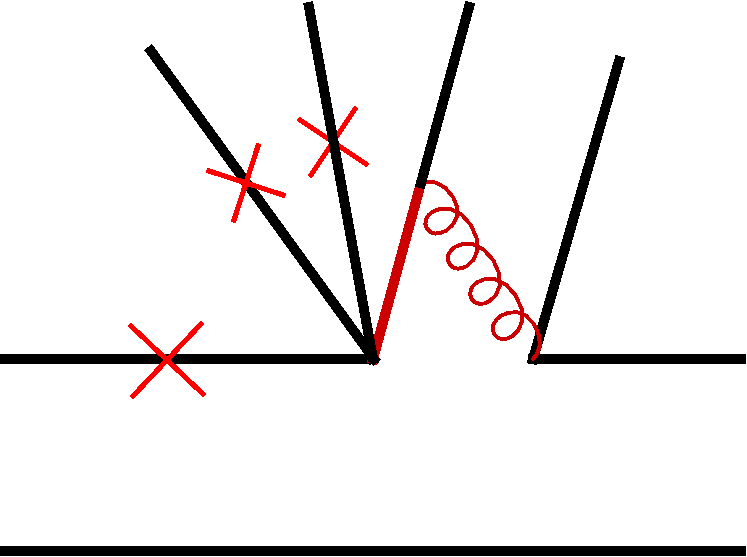}
\rput(-0.72,2.4){$\overbrace{\hspace{10mm}}^{\pi\pi}$}
\rput(0.5,0.32){$\Big\} {\scriptstyle \pi}$}
\end{pspicture}
\vspace{2mm}
\end{minipage}\\
\hline
&&&&&&\\
\rotatebox{90}{\hspace{-12mm} Edge (QCDF$_\text{II}$)}
&
\begin{minipage}{2.2cm}
\centering
\begin{pspicture}(0.8,2)(0,0)
\includegraphics[width=1.5cm,height=1.5cm]{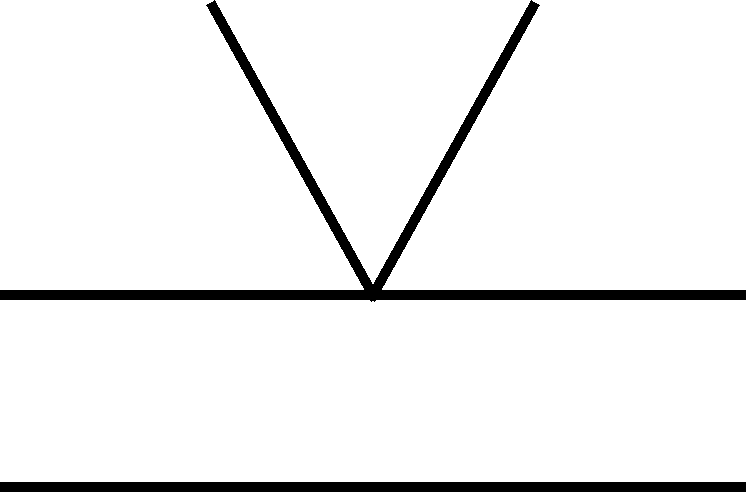}
\rput(-0.72,2.){$\overbrace{\hspace{10mm}}^{\phi_{\pi\pi}}$}
\rput(0.5,0.27){$\Big\} {\scriptstyle F_{\pi}}$}
\end{pspicture}\\
\vspace{2mm}
{\footnotesize Leading}
\vspace{4mm}
\end{minipage}
&
\begin{minipage}{2.3cm}
\centering
\begin{pspicture}(0.8,2)(0,0)
\includegraphics[width=1.5cm,height=1.5cm]{Corresp7.png}
\rput(-0.72,2.){$\overbrace{\hspace{10mm}}^{\phi_{\pi}}$}
\rput(0.5,0.27){$\Big\} {\scriptstyle F_{\pi\pi}}$}
\end{pspicture}\\
\vspace{2mm}
{\footnotesize Leading}
\vspace{4mm}
\end{minipage}
&
\begin{minipage}{2.3cm}
\centering
{\footnotesize ``Non-factorizable"}\\[2mm]
{\footnotesize Power-suppressed}
\vspace{3mm}
\end{minipage} &
\begin{minipage}{2.2cm}
\centering
{\footnotesize ``Non-factorizable"}\\[2mm]
{\footnotesize Power-suppressed}
\vspace{3mm}
\end{minipage} &
\begin{minipage}{2.4cm}
\centering
{\footnotesize 6-quark operator}\\[2mm]
{\footnotesize Power-suppressed}
\vspace{3mm}
\end{minipage} &
\begin{minipage}{2.2cm}
\centering
{\footnotesize 6-quark operator}\\[2mm]
{\footnotesize Power-suppressed}
\vspace{3mm}
\end{minipage}\\
\hline
\end{tabular}
\caption{Diagrammatic correspondence between the different contributions in QCDF$_\text{I}$ and QCDF$_\text{II}$.
Crosses denote alternative insertions of the gluon. One- and two-pion distributions are denoted by $\phi_\pi$ and $\phi_{\pi\pi}$
respectively, while $F_\pi$ and $F_{\pi\pi}$ denote $B\to \pi$ and $B\to\pi\pi$ form factors. The last four contributions
are leading at the center but power-suppressed at the edge. See the text for details.}
\label{tab:correspondence} 
\end{table}

We note at this point that in the center, since all invariant masses are large and of order $m_B^2$, there are always two
hard propagators, leading to an amplitude that is power suppressed with respect to the amplitude at the edge. In addition,
the perturbative nature of the hard gluon exchange leads to an $\alpha_s(m_b)$ suppression at the center, which is not
present at the edge, where the gluon becomes soft. All in all, the amplitude at the center is expected to be both
power- and $\alpha_s$-suppressed with respect to the amplitude at the edge.

\begin{figure}[t!]
\begin{center}
\includegraphics[width=8cm]{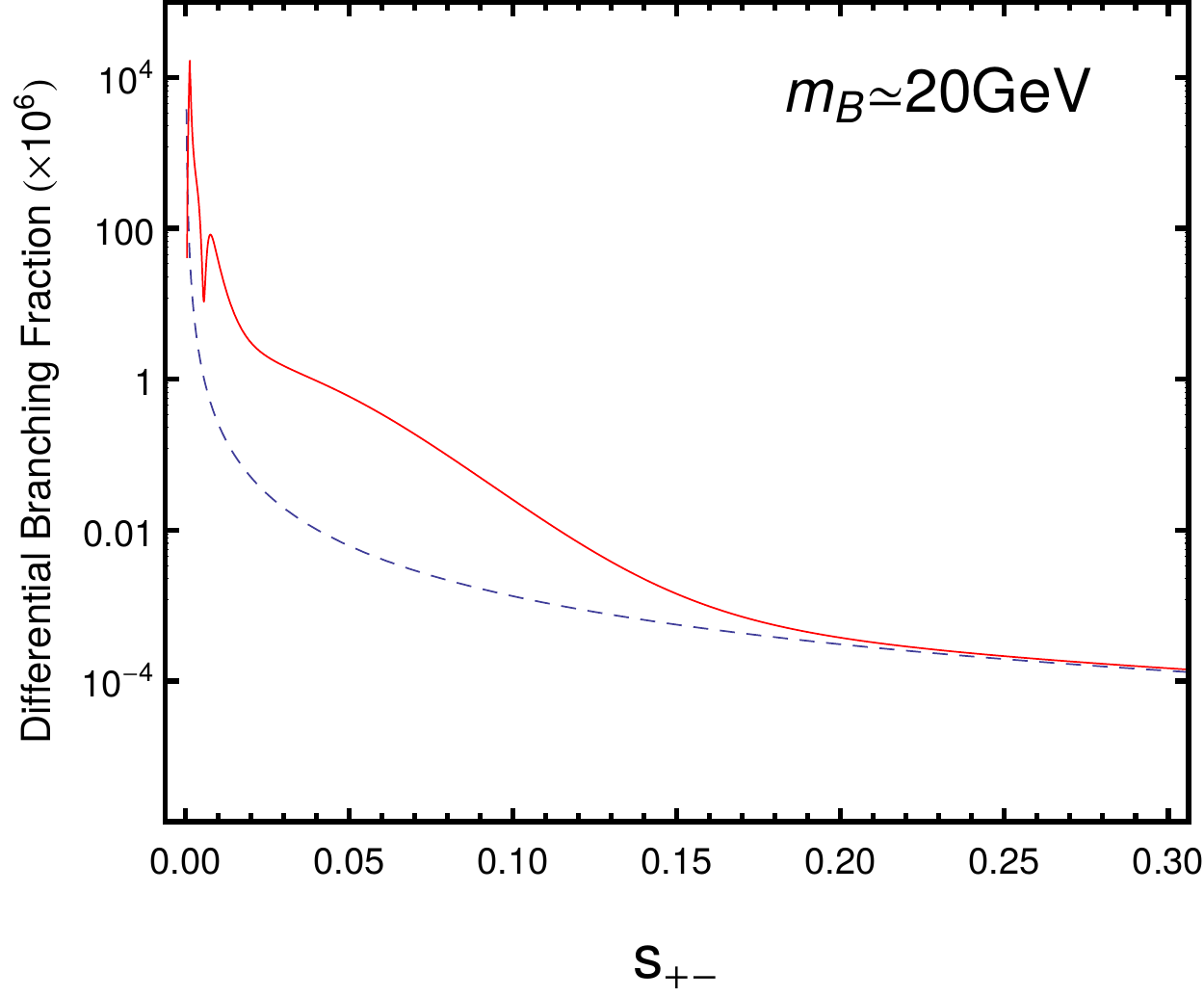}
\hspace{5mm}
\includegraphics[width=8cm]{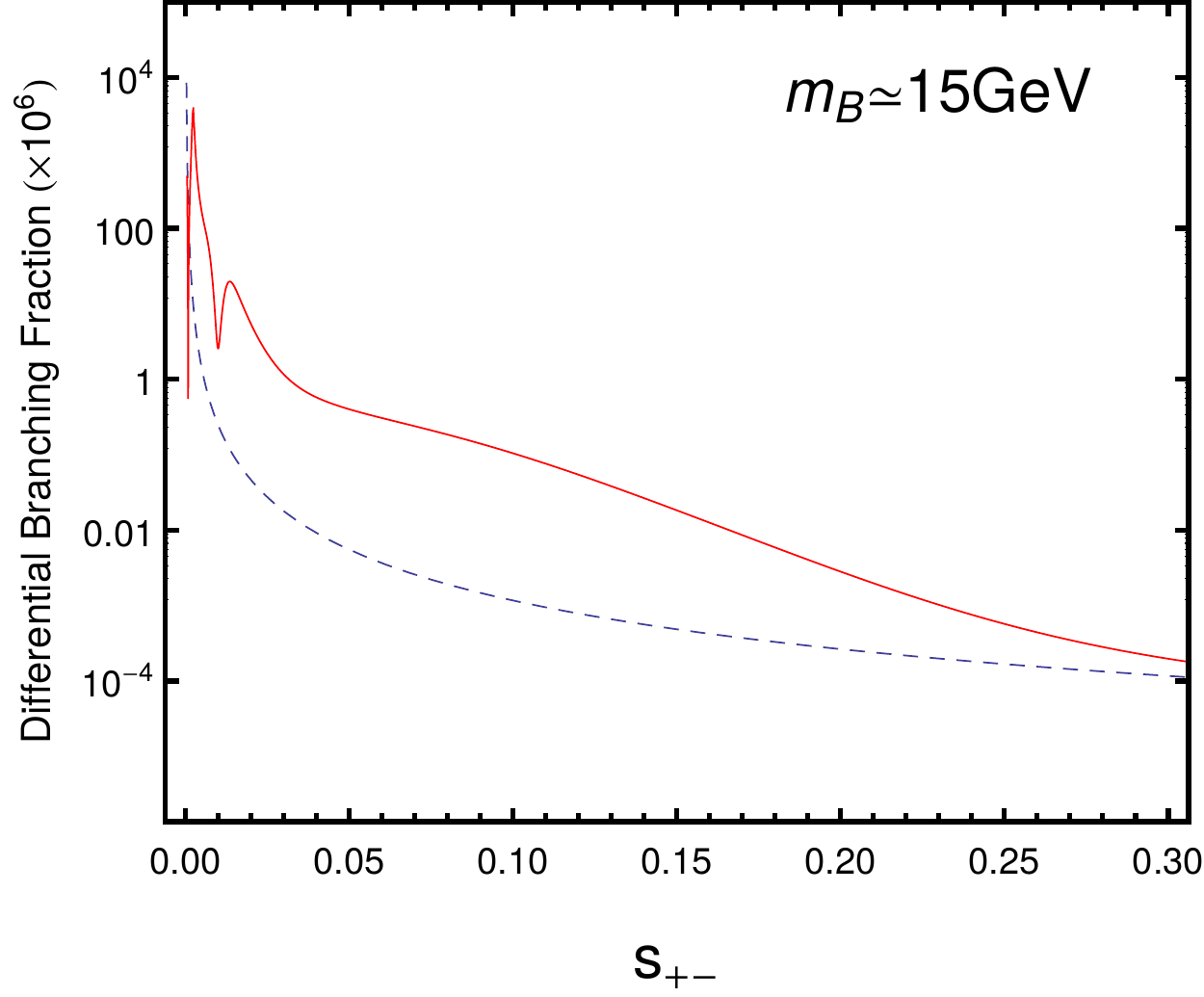}\\[5mm]
\includegraphics[width=8cm]{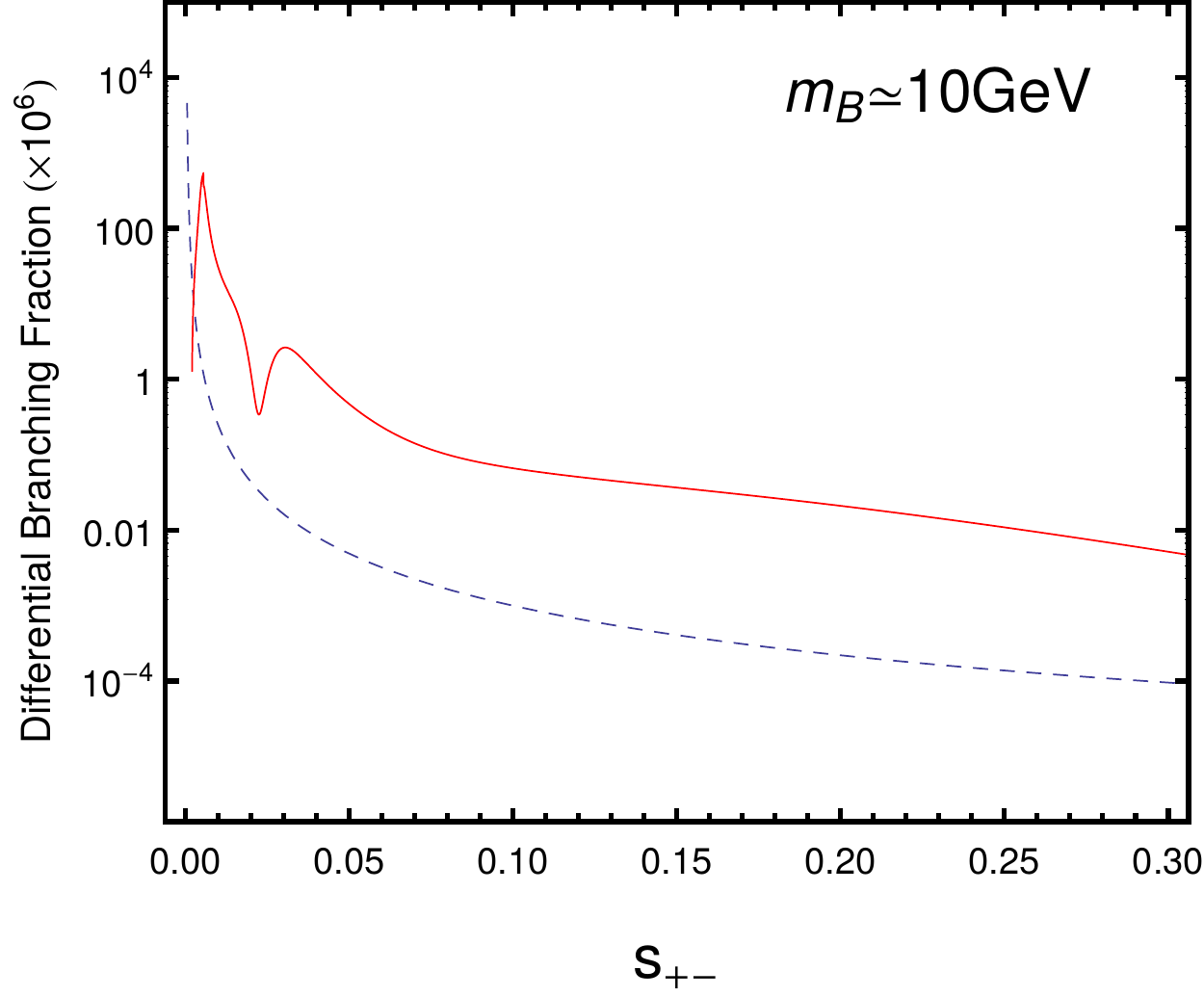}
\hspace{5mm}
\includegraphics[width=8cm]{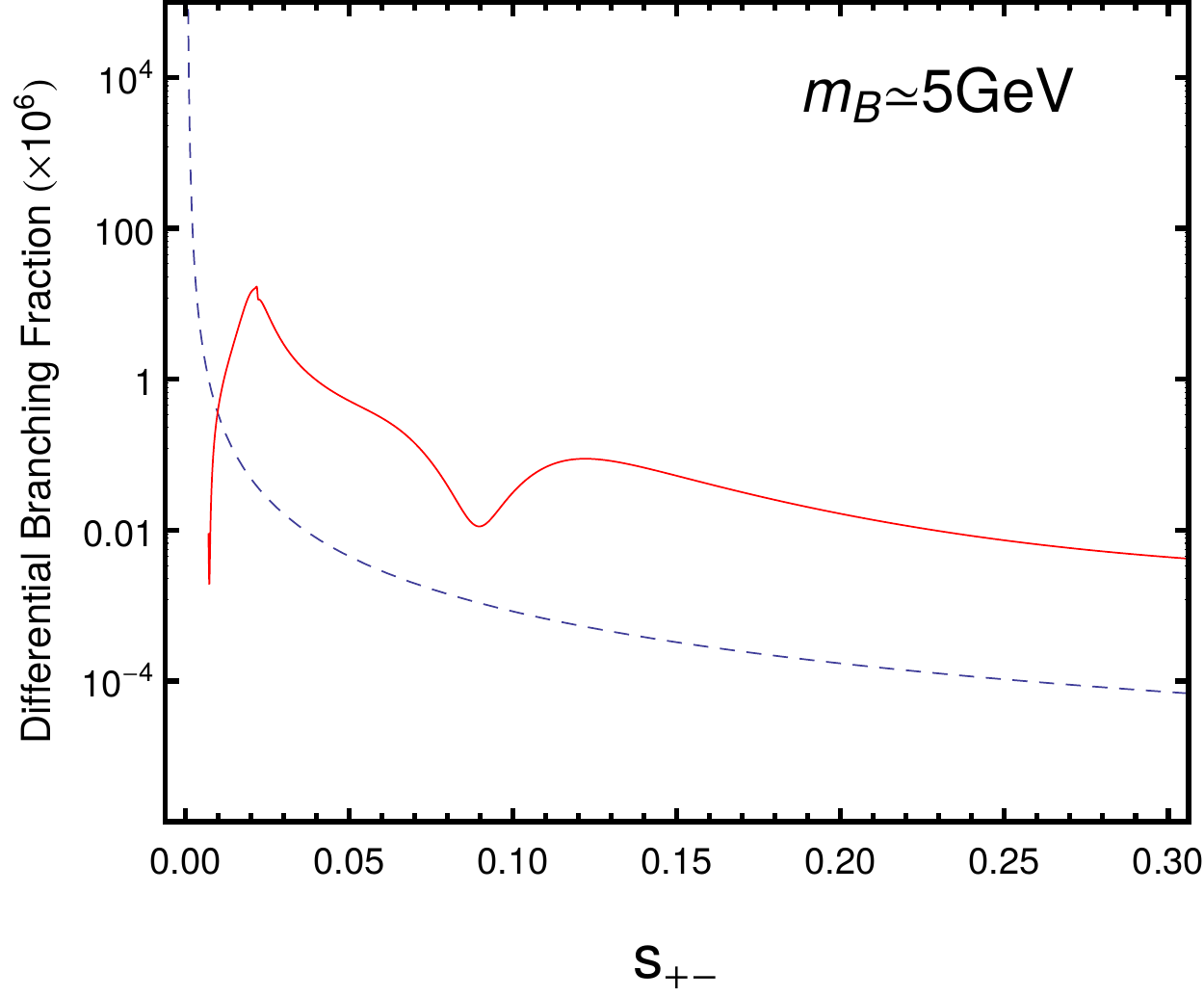}
\end{center}
\caption{Contributions from $2\pi$LCDs to the $B^+\to\pi^+\pi^-\pi^+$ differential branching fraction, for
$s_{++} = (1-s_{+-})/3$: Full contribution (solid) and perturbative contribution (dashed). A perturbative region
exists for large $s_{+-}$ in the heavy-quark limit, but probably not for realistic values of the $b$-quark mass.}
\label{fig:merge}
\end{figure}

While the previous considerations imply that formally there must be a good matching between both regions, the question is whether
this happens in practice for realistic $B$-meson masses. To this end we focus on the $2\pi$LCD contribution shown in the
first row in Table~\ref{tab:correspondence}. This contribution arises from the second term in Eq.~(\ref{eq:Aedge}). We find that,
in the limit of large $(m_B^2 s_{+-})$, this amplitude reproduces the corresponding contribution obtained from the
QCDF$_\text{I}$ calculation in Section~\ref{sec:MS} \footnote{This happens by construction, since we force the function $F_\pi(s)$ to satisfy
the perturbative limit asymptotically for large $s\,m_B^2$. This is in fact the only information we have on $F_\pi(s)$ at large energies,
since data reaches only up to $\sim 3$ GeV. For our purposes, the relevant observation is that data shows that the perturbative regime might lie beyond 3 GeV.}.
The particular values of $s_{+-}$ for which this matching occurs
depends on the value of $m_B^2$. In Fig.~\ref{fig:merge} we show the results of both calculations for different values
of $m_B$. We see that for $m_B\sim 20$ GeV there is enough phase space to reach a perturbative regime in the central region
of the Dalitz plot. However, the phase space gets reduced considerably when $m_B$ is decreased to its real value,
where there seems to be no perturbative regime, i.e. the Dalitz plot is completely dominated by the edges.

Similar conclusions are expected for the $B\to\pi\pi$ form-factor contribution in the second column in
Table~\ref{tab:correspondence}. Adding the rest of the central-region contributions to the perturbative side, we
will get a mismatch at large $(m_B^2 s_{+-})$ of the order of the perturbative contribution itself, which is
expected to be of the same order as neglected power corrections to the QCDF$_\text{II}$ calculation in the perturbative limit.
This corresponds to the contributions in the last four columns in Table~\ref{tab:correspondence}. Since we anyway do not
expect the two-pion system to factorize into single-pion distributions as early as $m_B^2 s_{+-}\sim 8$~GeV$^2$,
we can conclude that the QCDF$_\text{I}$ calculation of Section~\ref{sec:MS} might not be relevant in any region of
the Dalitz plot.

\section{Summary and conclusions}

Three-body $B$ decays provide many opportunities for studies of flavor physics and CP violation, as well for
studies of factorization issues in QCD. While a large amount of experimental information is already available,
an even larger amount is expected from future studies at the LHC, as well as from Belle-II. These promising
experimental prospects are not yet backed-up by theoretical studies able to describe these decays differentially
in the kinematics, and in a model-independent manner. In this paper we have performed a first study in the
context of QCD factorization.

The Dalitz plot for three-body $B$ decays can be divided in several regions with special kinematics and with different
factorization properties. In the heavy-quark limit, the amplitude in the central region of the Dalitz plot factorizes
into regular $B\to\pi$ form factors and pion light-cone distribution amplitudes. On the other hand, near the
edge two pions become collinear and the amplitude resembles a two-body decay, with the difference that
a new type of non-perturbative functions must be introduced: $B\to \pi\pi$ form factors and $2\pi$LCDs. The fact that
these objects cannot be factorized further is signaled by the divergence of the factorized expressions at the center
when one invariant mass is taken to zero.
We have calculated the amplitudes in both regions at the leading order, and verified these factorization properties. 

Assuming that these two regions are well described by the respective calculations, it is interesting to determine
how well these two descriptions merge at intermediate kinematical regimes. We have seen that some of the contributions
at the center correspond, in the heavy quark limit, to the expression for the amplitude at the edge with factorized
$B\to\pi\pi$ form factors and $2\pi$LCDs. Therefore, a parametrization of these nonperturbative objects that is
consistent with their perturbative limit leads automatically to a well behaved limit of the result at the edge
when extrapolated to the center. However, it seems that the perturbative regime is only kinematically allowed
for $b$-quark masses several times larger than the real value. In addition, the rest of the QCD factorization contributions
at the center correspond to power-suppressed corrections at the edge.

At this point, a more refined study of the Dalitz plot distribution based on a factorization in terms of
two-pion distributions and $B\to\pi\pi$ form factors seems worthwhile. Next-to-leading (NLO) corrections to the hard kernels
are already known from two-body decays, and can be used directly in the factorization formula of Eq.~(\ref{s23small}) to verify
factorization at NLO. On the phenomenological level, this requires a better knowledge of two-pion distributions. First, one must go beyond the
local limit (where information other than $F_\pi(s)$ is needed). Second, the tensor distribution $\Phi_\bot$ is
expected to contribute, while little is known about it. Besides the traditional studies of two-pion distributions from
$\gamma^*\gamma\to \pi\pi$ and $\tau\to\pi\pi \nu_\tau$, we propose to study such distributions in the context of
non-leptonic decays such as $B^0_s\to D_s^-\pi^+\pi^0$ where factorization is considerably simpler than for $B\to \pi\pi\pi$.
Regarding $B\to\pi\pi$ form factors, better knowledge is also required. Improved light-cone sum-rule calculations
\cite{HK}, as well as precise experimental studies of the semileptonic decays $B\to\pi\pi\ell\nu$ \cite{1310.6660}
and $B\to\pi\pi\ell\ell$ will be essential.

Besides $B\to\pi\pi\pi$ decays, other three-body decays with kaons ($B^+\to K^+ \pi^-\pi^+$, etc.) have been studied
experimentally at $B$-factories and the LHC (e.g. \cite{1310.4740,1501.00705}). Their branching fractions are higher because they are not CKM suppressed,
with the corresponding impact in terms of statistics. These channels can be studied in a similar fashion.
This requires knowledge on $B\to K\pi$ and $B\to KK$ form factors, as well as $K \pi$ and $KK$
distributions. Again, these can be accessed from semileptonic $B$ decays (e.g. \cite{1406.6681}) and
$\tau$ decays (e.g. \cite{1007.1858}).

\section*{Acknowledgements}

We thank Martin Beneke for very useful discussions and for sharing with us his notes on the subject \cite{talkMB}.
We thank Tobias Huber for collaboration at early stages of this work and for extensive discussions.
We thank Thorsten Feldmann, Christian Hambrock, Alex Khodjamirian, Bjorn Lange , Danny van Dyk, Pablo Roig, Rafel Escribano and
Pere Masjuan for discussions and correspondence.
We thank Marc Grabalosa and Fernando Rodrigues for correspondence in relation with the LHCb analysis of
Ref.~\cite{1408.5373} and Tim Gershon, John Back and Gagan Mohanty concerning the BaBar results of Ref.~\cite{0902.2051}.
We also acknowledge useful discussions with Luca Silvestrini and Yuming Wang at the B2TiP Workshop in Krakow.
This work has been funded by the Deutsche Forschungsgemeinschaft (DFG) within research unit FOR 1873 (QFET).


\end{document}